\documentclass[aps,prl,floatfix,showpacs,twocolumn,preprintnumbers,tightenlines,superscriptaddress]{revtex4-1}

\usepackage{papersstyle}

\newcommand{\TOne}{\ensuremath{T_{\rm 1}}}
\newcommand{\TEcho}{\ensuremath{T_{\rm 2}^{\rm Echo}}}

\newcommand{\Ts}{\ensuremath{T_{\rm 2}^*}}
\newcommand{\GEcho}{\ensuremath{\Gamma_{\rm \varphi}^{\rm Echo}}}
\newcommand{\GEchomin}{\ensuremath{\Gamma_{\rm{\varphi, min}}^{\rm Echo}}}
\newcommand{\fR}{\ensuremath{f_{\rm R}}}
\newcommand{\fbare}{\ensuremath{f_{\rm bare}}}
\newcommand{\deltar}{\ensuremath{\Delta f_{\rm R}}}
\newcommand{\fQ}{\ensuremath{f_{01}}}
\newcommand{\fOtwoH}{\ensuremath{f_{\rm 02}/2}}
\newcommand{\Deltaf}{\ensuremath{\Delta f}}
\newcommand{\Bpar}{\ensuremath{B_\parallel}}
\newcommand{\Bc}{\ensuremath{B_{\rm c}}}
\newcommand{\Qdiel}{\ensuremath{Q_{\rm{d}}}}
\newcommand{\Vg}{\ensuremath{V_{\rm G}}}
\newcommand{\phisqrd}{\ensuremath{\langle\phi^2(t)\rangle}}
\newcommand{\SqrtAV}{\ensuremath{\sqrt{A_{\rm{V}}}}}
\newcommand{\EJ}{\ensuremath{E_{\rm J}}}
\newcommand{\EC}{\ensuremath{E_{\rm C}}}
\newcommand{\Qi}{Q_{\mathrm{i}}}
\newcommand{\Qc}{Q_{\mathrm{c}}}
\newcommand{\Tesla}{\ensuremath{\mathrm{T}}}
\newcommand{\mT}{\ensuremath{\mathrm{mT}}}
\newcommand{\s}{\ensuremath{\mathrm{s}}}
\newcommand{\us}{\ensuremath{\mu\mathrm{s}}}
\newcommand{\ns}{\ensuremath{\mathrm{ns}}}
\newcommand{\ms}{\ensuremath{\mathrm{ms}}}
\newcommand{\persec}{\ensuremath{\mathrm{s}^{-1}}}
\newcommand{\uFlux}{\ensuremath{\mu\Phi_0}}
\newcommand{\nFlux}{\ensuremath{\mathrm{n}\Phi_0}}
\newcommand{\Hz}{\ensuremath{\mathrm{Hz}}}
\newcommand{\kHz}{\ensuremath{\mathrm{kHz}}}
\newcommand{\MHz}{\ensuremath{\mathrm{MHz}}}
\newcommand{\GHz}{\ensuremath{\mathrm{GHz}}}
\newcommand{\uV}{\ensuremath{\mu\mathrm{V}}}
\newcommand{\nm}{\ensuremath{\mathrm{nm}}}
\newcommand{\um}{\ensuremath{\mu \mathrm{m}}}
\newcommand{\V}{\ensuremath{\mathrm{V}}}
\newcommand{\mK}{\ensuremath{\mathrm{mK}}}
\newcommand{\Cel}{\ensuremath{^\circ\mathrm{C}}}
\newcommand{\mA}{\ensuremath{\mathrm{mA}}}
\newcommand{\kHzpersqrtHz}{\ensuremath{\mathrm{kHz}/\sqrt{\mathrm{Hz}}}}

\begin{document}

\title{Evolution of Nanowire Transmons and Their Quantum Coherence in Magnetic Field}

\author{F.~Luthi}
\author{T.~Stavenga}
\author{O.~W.~Enzing}
\author{A.~Bruno}
\author{C.~Dickel}
\author{N.~K.~Langford}
\author{M.~A.~Rol}
\affiliation{QuTech, Delft University of Technology, Delft, The Netherlands}
\affiliation{Kavli Institute of Nanoscience, Delft University of Technology, Lorentzweg 1, 2628 CJ Delft, The Netherlands}
\author{T.~S.~Jespersen}
\affiliation{Center for Quantum Devices, Niels Bohr Institute, University of Copenhagen, Copenhagen, Denmark}
\author{J.~Nyg\r{a}rd} 
\affiliation{Center for Quantum Devices, Niels Bohr Institute, University of Copenhagen, Copenhagen, Denmark}
\affiliation{Nano-Science Center, Niels Bohr Institute, University of Copenhagen, Copenhagen, Denmark}
\author{P.~Krogstrup}
\affiliation{Center for Quantum Devices, Niels Bohr Institute, University of Copenhagen, Copenhagen, Denmark}
\author{L.~DiCarlo}
\affiliation{QuTech, Delft University of Technology, Delft, The Netherlands}
\affiliation{Kavli Institute of Nanoscience, Delft University of Technology, Lorentzweg 1, 2628 CJ Delft, The Netherlands}

\begin{abstract}

We present an experimental study of nanowire transmons at zero and applied in-plane magnetic field.
With Josephson non-linearities provided by the nanowires, our qubits operate at higher magnetic fields than standard transmons. 
Nanowire transmons exhibit coherence up to $70~\mT$, where the induced superconducting gap in the nanowire closes.
We demonstrate that on-chip charge noise coupling to the Josephson energy plays a dominant role in the qubit dephasing.
This takes the form of strongly-coupled two-level systems switching on $100~\ms$ timescales and a more weakly coupled background producing $1/f$ noise. 
Several observations, including the field dependence of qubit energy relaxation and dephasing, are not fully understood, inviting further experimental investigation and theory. 
Using nanowires with a thinner superconducting shell will enable operation of these circuits up to $0.5~\Tesla$, a regime relevant for topological quantum computation.

\end{abstract}

\maketitle

Circuit quantum electrodynamics (cQED) offers unprecedented control over coupled atomic and photonic degrees of freedom in engineerable, microscale superconducting circuits~\cite{Blais04, Wallraff04}.
It relies crucially on the dissipationless nonlinearity of the Josephson effect between two weakly coupled superconductors~\cite{Josephson1962}.
The Josephson junction (JJ), usually implemented in the form of superconductor--insulator--superconductor (SIS) junctions, allows the realization of anharmonic oscillators that can be operated in the quantum regime and used as qubits~\cite{Nakamura99}.
Circuit QED has found applications in many areas, including scaleable quantum computation~\cite{Barends14}, quantum optics~\cite{Steffen13}, quantum foundations~\cite{Jerger16} and quantum measurement and control~\cite{Riste13c}.
So far, cQED has been limited by standard SIS JJs based on aluminum and its oxide to fields $<10~\mT$, the critical field of bulk aluminum~\cite{Harris1968}.
However, interesting applications such as coupling cQED devices to polarized electron-spin ensembles serving as quantum memories~\cite{Imamoglu09} and using qubits as charge-parity detectors in Majorana based topological quantum computation~\cite{Hyart13, Mourik12} require fields of $\sim0.5~\Tesla$.
In such fields, more fundamental effects such as topological phase transitions~\cite{Kosterlitz1973} and degeneracy-lifting of the Andreev bound states which underlie the Josephson effect~\cite{Andreev1964,Pillet10,vanWoerkom17,Yokoyama13} can be studied.
Entering this important regime for cQED requires the use of field-compatible superconductors and non-standard JJs~\cite{Samkharadze16, Popinciuc12, Doh05, Pallecchi08, Della07}.

To date, qubits in cQED architectures have been realized using various JJs: the ubiquitous SIS tunnel junction~\cite{Nakamura99}, atomic break junctions~\cite{Janvier15} and semiconductor weak-link nanowire junctions~\cite{deLange15,Larsen15, Casparis15}. 
Nanowire qubits are of particular interest because of potential high-magnetic field compatibility, the voltage tunability of the JJ and the overlap with other technologies of interest, including nanowire-based transistors and lasers~\cite{Chuang13, Liu15}. 
Nanowire qubits are compatible with the transmon geometry~\cite{Koch07}, the most widely used in cQED, and have been realized in flux and voltage tunable variants~\cite{Larsen15, deLange15}.
Nanowire transmons have reached echo dephasing times ($\TEcho$) up to 10~$\us$, and been used to implement two-qubit gates~\cite{Casparis15}.
So far, the use of Al as a superconductor for the larger scale cQED elements~\cite{Larsen15,Casparis15} and short coherence times~\cite{deLange15} have inhibited study of the coherence of these circuits in a magnetic field.

In this Letter, we characterize noise processes affecting nanowire transmons and explore their behavior in an in-plane magnetic field ($\Bpar$).
Using a flux-tunable split-junction device, we first achieve $\TEcho$ limited by energy relaxation ($\TOne$) at the flux sweet spot.
Independent of the flux-noise reducing coherence away from the sweet spot, the qubit coherence suffers from a charge two-level system (TLS) strongly coupled to the nanowire Josephson energy $\EJ$, leading to a switching of the qubit frequency ($\fQ$).
The TLS switching is observed in real time using a single-shot, frequency-detecting pulse sequence giving characteristic switching times of 100~$\ms$~\cite{Riste13}.
We also observe this frequency switching in side-gate-tunable single-junction qubits (gatemons) and show that it is sensitive to the applied side-gate voltage $\Vg$.
In addition to these strongly coupled TLSs, a weakly coupled $1/f$ charge-noise background is also investigated.
Finally, we study the behavior of a gatemon in applied $\Bpar$, observing reduced $\fQ$ as the superconducting gap ($\Delta$) induced in the nanowire is suppressed.
While both $\TOne$ and $\TEcho$ are reduced due to $\Bpar$, an effect not yet understood, the device exhibits measurable coherence up to $70~\mT$.

The device fabrication combines procedures widely used in cQED with nanowire etching and contacting recipes.
Large features on the chips (ground plane, waveguides and transmon capacitors) are defined by a reactive-ion etch of the NbTiN film on the high-resistivity silicon substrate~\cite{Thoen17,Bos17,Bruno15}.
The nanowires used to create the superconductor-semiconductor-superconductor (SNS) junction have an InAs core and an epitaxially grown Al shell that induces a hard superconducting gap, suppressing quasiparticle tunneling~\cite{Doh05, Chang15, Martinis09, Wang14}.
They are deterministically placed in the contacting region using a nanomanipulator.
The individual positions of the nanowires are analyzed using a home-developed image recognition software that automatically defines etch- and contacting masks~\cite{SM, opencv}.
After defining the SNS junction by wet-etching a 200~$\nm$ segment of the 30~$\nm$ thick Al shell, the wires are contacted with NbTiN.
The qubits are individually coupled to dedicated readout resonators which each couple to a common feedline, allowing for multiplexed readout~\cite{Jerger12}.
Standard cQED control and measurements schemes are used to probe the qubits~\cite{Blais04, Motzoi09}.

\begin{figure}[t]
  \centering
  \includegraphics[width=\linewidth]{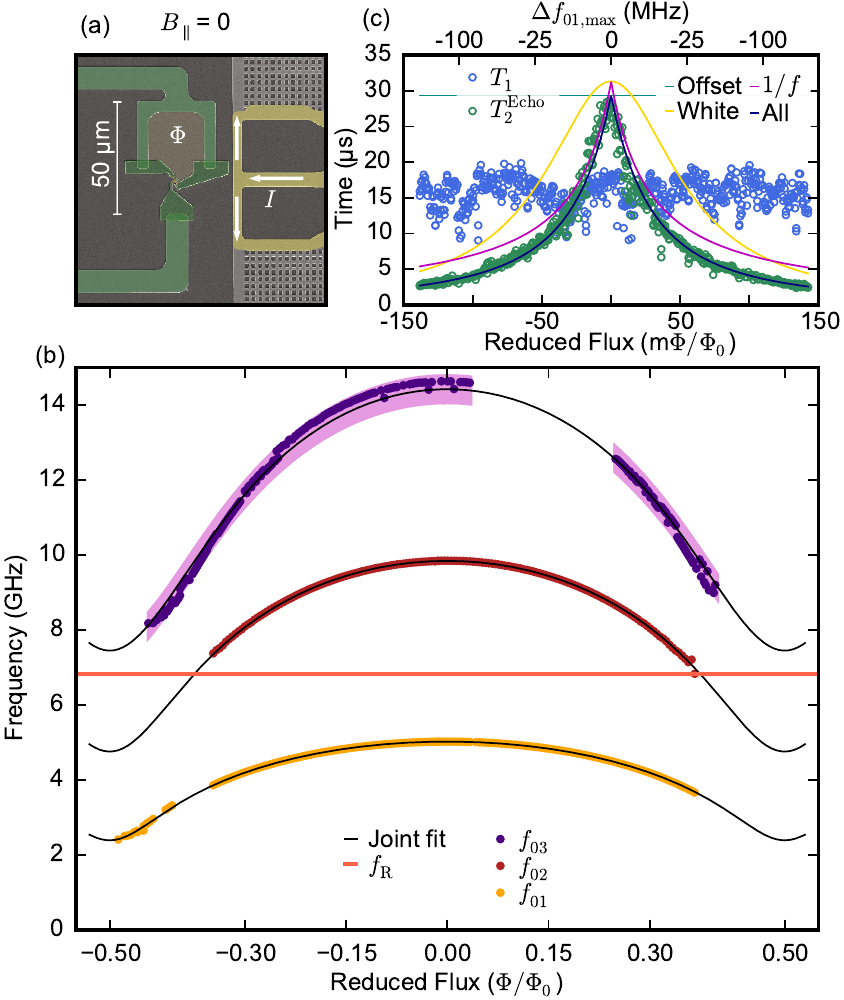}
  \caption{
Characterization of a flux-tunable split-junction qubit at $\Bpar=0~\mT$.
(a) False-colored SEM micrograph of the SQUID-loop area. The current $I$ in the flux-bias line (yellow) threads a net magnetic flux $\Phi$ through the asymmetrically positioned SQUID-loop of the transmon (green), tuning $\EJ(\Phi)$. 
(b) The joint fit (black) of the spectroscopy versus applied flux of the first three transitions (orange, dark red and purple symbols) yields the transmission probability and the induced gap of each junction. 
(c) $\TOne$ (blue) is limiting $\TEcho$ (green) at the qubit flux sweet spot $\Phi=0$. 
A fit to $\TEcho$ that includes the measured $\TOne$-limit allows to extract a flux-independent (cyan), a $1/f$ noise (pink) and white-noise (gold) contribution to the dephasing.
$\Ts$ is typically below $4~\us$. 
Top axis indicates the frequency detuning from the flux sweet spot.
}
  \label{fig:Fig1}
\end{figure}

We start by studying the spectrum of the flux-tunable, split-junction device, to extract information about the SNS junctions.
Applying a current $I$ through the flux-bias line changes the magnetic flux $\Phi$ through the SQUID loop [Fig.~\ref{fig:Fig1}(a)] and thus controls the superconducting phase difference $\hat\delta$ between the two transmon islands.
This phase difference tunes $\EJ$ in each junction, given in the short-junction, single-channel limit by Andreev bound states with transmission probability $T_i$ and energy $\V_i(\phi_i)=-\Delta_i\sqrt{1-T_i\sin^2(\phi_i/2)}$.
Employing the Andreev bound state model in the split-junction Cooper-pair box type Hamiltonian, $H=4\EC\hat N^2+V_A(\hat\delta)+V_B(2\pi\Phi/\Phi_0-\hat\delta)$, yields good agreement with the observed spectrum [Fig.~\ref{fig:Fig1}(b)], as in~\cite{deLange15}.
The best-fit values of the induced gaps $\Delta_A/h=46\pm 4~\GHz$ and $\Delta_B/h=38.5\pm0.9~\GHz$ are close to the bulk Al gap of 43~$\GHz$.
This shows that the Al shell fully proximitizes the nanowire~\cite{Wang14}.

We investigate the flux noise of the split-junction qubit by measuring coherence times as a function of flux offset.
The $\TEcho$ is $\TOne$ limited at the flux sweet spot, but is reduced as the sensitivity to flux noise increases [Fig.~\ref{fig:Fig1}(c)].
The noise is quantified using a quadratic fit to the echo dephasing rate $\GEcho=1/\TEcho-1/(2\TOne)$ plotted against $|\partial \fQ/\partial \Phi|$~\cite{Martinis03, Yoshihara06, Hutchings17, SM}.
A least-squares fit yields a white-noise contribution $S_{\Phi \mathrm{ , white}}=(60~\nFlux/\sqrt{\Hz})^2$, a $1/f$ noise amplitude $\sqrt{A_\Phi}=13.0~\uFlux$ where $S_{\Phi, \mathrm{1/f}}=A_\Phi/|f|$ and a flux-independent offset of 2~$\ms^{-1}$.
The small size of the flux-independent offset demonstrates that the transport through the nanowire junction is highly coherent.
The value of the $1/f$ flux noise amplitude is a factor of $\sim 10$ larger than that of SIS transmons~\cite{Hutchings17}.
The origin of the observed white noise contribution that is typically absent in flux-tunable transmon qubits is unclear.

\begin{figure}[t]
  \centering
  \includegraphics[width=\linewidth]{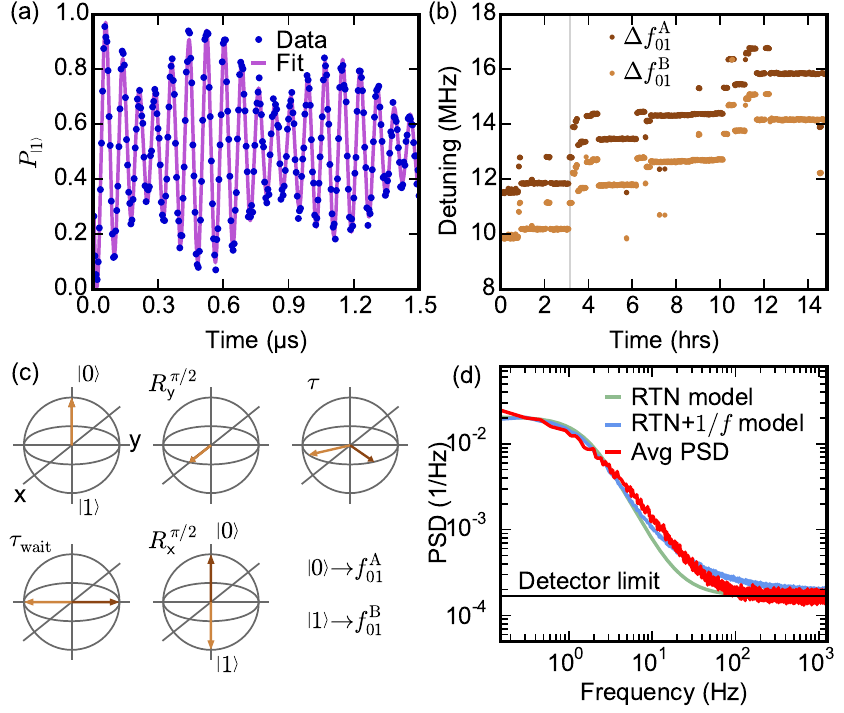}
  \caption{
Frequency stability analysis of the flux-tunable qubit at the flux sweet spot and at $\Bpar=0~\mT$.
(a) A Ramsey experiment (dots) with artificial detuning ($12~\MHz$). The strongly coupled TLS leads to a beating between two decaying sinusoids (purple, $\tau^\mathrm{A}=2.2~\us$ and $\tau^\mathrm{B}=2.0~\us$).
(b) The extracted detunings of repeated Ramsey experiments show a constant frequency spacing and drift of the center frequency. 
The gray vertical line indicates the trace shown in (a).
(c) Pulse sequence of the Ramsey-type TLS-state detection scheme. The free evolution time $\tau$ is chosen as $\tau_\mathrm{wait}=1/2\Deltaf$ for maximal contrast.
(d) The PSD (red) of the TLS is computed from traces of qubit states obtained by monitoring the qubit frequency real-time using the pulse sequence described in (c).
The PSD is fitted using RTN models with (blue) and without (green) $1/f$ noise.
}
  \label{fig:Fig2}
\end{figure}

Ramsey measurements reveal a switching of $\fQ$ from a strong coupling of charge TLSs to $\EJ$ of the nanowire, yielding a beating pattern of two exponentially decaying sinusoids [Fig.~\ref{fig:Fig2}(a)].
The observed frequency difference $\Deltaf=f_{01}^\mathrm{A}-f_{01}^\mathrm{B}=1.6~\MHz$ is consistent with the strong coupling of a TLS to the qubit ($1/\Deltaf<\tau^\mathrm{A},\tau^\mathrm{B}$, the respective dephasing times).
Repeating this measurement over 14~hours confirms a constant spacing between $\fQ^\mathrm{A, B}$ [Fig.~\ref{fig:Fig2}(b)] with an additional drift indicating further TLSs that switch on a much slower timescale.
Because $\Deltaf$ is much larger than the transmon's charge dispersion~\cite{Koch07} of only $200~\kHz$ and constant over time, in contrast to~\cite{Riste13}, we conclude that the TLS couples to the nanowire $\EJ$.
This strong TLS coupling is the reason why we do not report Ramsey coherence times ($\Ts$) in Fig.~\ref{fig:Fig1}.
The qubit can be used to measure the state of the TLS in real time using a single-shot Ramsey-based pulse sequence tailored for $\Deltaf$ [Fig.~\ref{fig:Fig2}(c)]~\cite{Riste13}.
Using the time evolution of the TLS state, we show that its double-sided power spectral density (PSD) can be explained by an asymmetric random telegraph noise (RTN) with characteristic switching times of $100~\ms$ [Fig.~\ref{fig:Fig2}(d)], see Ref.~\onlinecite{SM} for details.
Better agreement with the measured PSD is achieved by taking $1/f$ noise into account.
The switching of $f_{01}$ between multiple values can be observed in several qubits.
In addition, the frequency difference was observed to depend on the electrostatic environment of the junction for gatemons~\cite{SM}.
This dependence indicates that the TLSs are charge traps in the vicinity of the junction that influence the transmission probabilities of the Andreev bound states.

\begin{figure}[t]
  \centering
  \includegraphics[width=\linewidth]{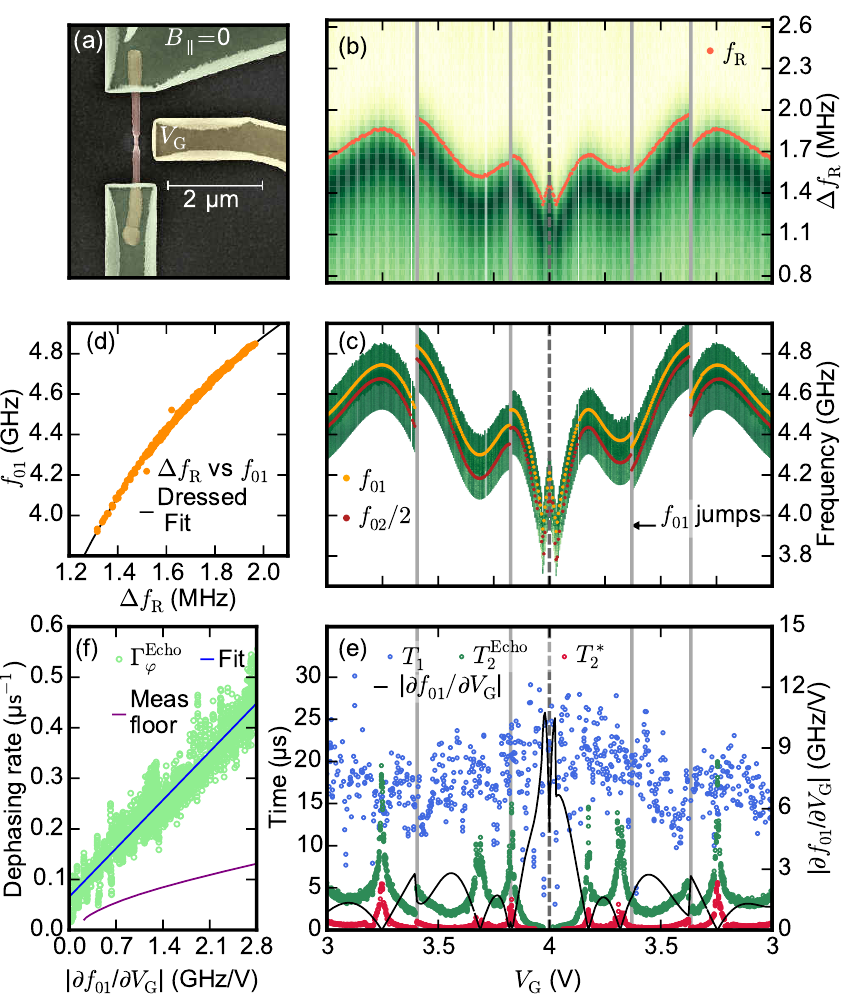}
  \caption{
Characterization of a gatemon at $B_\parallel=0~\mT$. 
(a) False-colored SEM micrograph of the nanowire Josephson junction (light red) with a side gate (yellow) that allows for $\Vg$ tuneability.
(b) Deviation of $\fR$, $\deltar$, from the bare resonator frequency $\fbare = 6.732~\GHz$ for a triangle sweep in $\Vg$. Note the change in direction of the $\Vg$ sweep, indicated by the dashed line. On return to the same $\Vg$, $\fR$ is roughly reproduced.
(c) $\fQ$ versus $\Vg$. Random, but sometimes reproducible jumps of $\fQ$ occur and are indicated by light gray lines.
(d) Plot of $\fQ$ against $\fR$ (orange dots) and dressed state fit (black curve) with coupling strength $g/2\pi = 60.8~\MHz$. The fit allows a quick and reliable prediction of $\fQ$. 
(e) Gatemon $\TOne$ (blue symbols), $\TEcho$ (green symbols) and $T_2^*$ (red symbols) versus $\Vg$. Both $\TEcho$ and $T_2^*$ are strongly correlated with the $\Vg$ sensitivity (black).
(f) $\Gamma_\varphi^\mathrm{Echo}$ against $\Vg$ sensitivity, extracted from (e). The fitted $1/f$ noise (blue) is clearly above the dephasing limit imposed by the setup noise (purple), indicating additional on-chip noise.
}
  \label{fig:Fig3}
\end{figure}

We now study the spectrum of a gatemon as a function of $\Vg$ [Fig.~\ref{fig:Fig3}(a-d)].
The $\Vg$ tuning changes $\fQ$ by altering the $T_i$ of the Andreev bound states, hence altering $\EJ$.
The tuning is repeatable upon small excursions (1-2~$\V$), except for isolated deviations which we attribute to charge traps changing state.
These changes --- some are reproducible, others are stochastic --- lead to jumps in $\fQ$.
Because the gatemon-resonator pair is well described by the dressed-state picture~\cite{Blais04}, $\fQ$ is easily found after a jump by measuring $\fR$ and calculating the expected $\fQ$.

For the gatemon, the dominant source of dephasing is on-chip $1/f$ voltage noise.
A clear correlation between the $\Vg$ sensitivity of the qubit, $|\partial \fQ/\partial \Vg|$, and $\Ts$ and $\TEcho$ is observed [Fig.~\ref{fig:Fig3}(e)].
The ratio between echo and Ramsey coherence times of roughly 8 (data not shown) indicates that the white noise contribution is small~\cite{Martinis03}.
Therefore, only a linear fit to $\GEcho$ plotted against $|\partial \fQ/\partial \Vg|$ is performed [Fig.~\ref{fig:Fig3}(f)]~\cite{Martinis03, Yoshihara06, Hutchings17, SM}.
The extracted quantities are a voltage-noise-independent offset of $66~\ms^{-1}$ and a $1/f$ voltage noise amplitude $\sqrt{A_{\mathrm{V}}}=26~\uV$, where $S_{\mathrm{V, 1/f}}=A_\mathrm{V}/|f|$.
The extracted noise is clearly larger than the noise floor of the biasing circuit~\cite{SM}, indicating charge noise on the chip as the dominant source of noise.
Possible origins of the dominant charge noise are the substrate or surface absorbents~\cite{SM, Gul15}.

\begin{figure*}[t]
  \centering
  \includegraphics[width=\linewidth]{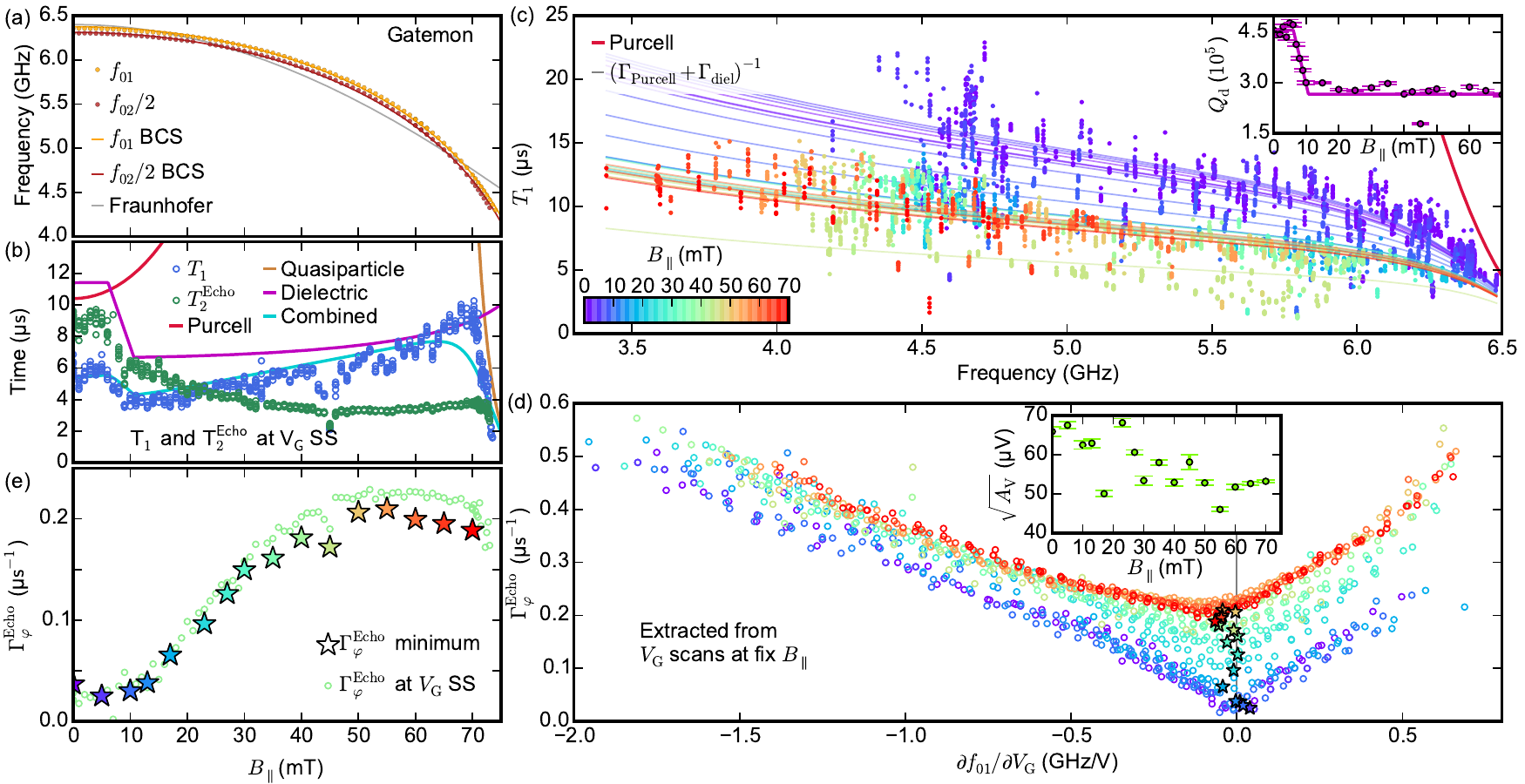}
  \caption{
  Evolution of the gatemon of Fig.~\ref{fig:Fig3} in $\Bpar$. 
(a) Qubit $\fQ$ and $\fOtwoH$ (orange and red) are well described by a closing BCS gap with $\Bc=83.9\,$mT (curves, see text for the model). 
(b) At each value of $\Bpar$, the gatemon is tuned to a sweet spot in $\Vg$ to measure $\TOne$ and $\TEcho$ (blue and green points).
At low $\Bpar$ ($\fQ$ near the resonator), $\TOne$ is mainly limited by the Purcell effect (red).
At $\Bpar$ close to $\Bc$ the superconducting gap becomes so weak that $\TOne$ is limited by quasiparticles (brown).
In-between, the $\TOne$ evolution can be attributed to a changing $\Qdiel$.
(c) $\TOne$ versus frequency at different $\Bpar$ ($\Bpar$ increases along the rainbow spectrum).
Accounting for $\TOne^{\rm{P}}$, we fit a $\Qdiel$ at each $\Bpar$ (inset), revealing a drop above $6~\mT$.
(d) Keeping $\Bpar$ fixed [same color scale as (c)], scans in $\Vg$ are performed to extract $\GEcho$, similar to Fig.~\ref{fig:Fig3}(e).
Because the dephasing rate minima (stars) do not coincide with the $\Vg$ sweet spot and the slopes of $\GEcho$ are not symmetric, we surmise that the leverarms of the side gate and the dominant noise source are comparable.
Inset: averaged extracted $1/f$ voltage noise amplitude.
(e) Pure dephasing rates at $\Vg$ sweet spots versus $\Bpar$ from data in (b) and (d). 
Stars denote the interpolated minimal dephasing rates from (d).
   }
  \label{fig:Fig4}
\end{figure*}

We now apply a $\Bpar$ to the same gatemon.
We focus on the gatemon because flux-tunable devices experience fluctuating $\fQ$ due to imperfect alignment and limited stability of $\Bpar$.
In order to disentangle $\Bpar$ and $\Vg$ contributions, the gatemon is placed at the same $\Vg$ sweet spot for each $\Bpar$ value.
We directly attribute the observed monotonic decrease in $\fQ$ with $\Bpar$ [Fig.~\ref{fig:Fig4}(a)] to a reduction of the induced superconducting gap in the nanowire junction, $\Delta (\Bpar)=\Delta(0)(1-(\Bpar/\Bc)^2)^{1/2}$~\cite{Tinkham96}.
We expect the bulk of the cQED chip to exhibit little change due to the high parallel critical field ($\Bc$) of the NbTiN film.
We approximate the energy of the Andreev bound state with $V_i(\phi_i,\Bpar)=-\Delta(\Bpar)\sqrt{1-T_i\sin^2(\phi_i/2)}$.
The Hamiltonian $H=4\EC\hat N^2+V_A(\hat\delta,\Bpar)+V_B(\hat\delta,\Bpar)$ is then fitted to $\fQ$ and $\fOtwoH$, fixing $\Delta(0)$ to the bulk Al gap and $\EC$ to the value obtained for the split-junction device.
The best-fit agrees well with the data and yields the free parameters $T_{A,B} = 0.95,\,0.62$ and $\Bc = 83.9~\mT$.
The extracted $\Bc$ of the Al shell is in good agreement with other measurements of wires from this growth batch~\cite{Chang15}.
Because $\Bpar$ is not collinear with the nanowires, the $\Bc$ of different qubits varies between $50$ and 90~$\mT$, with a rough correlation to the angle between the nanowire and the direction of $\Bpar$~\cite{SM}.
The $\Bc$ values stay constant over the duration of a cooldown and vary by $\sim5~\%$ between cooldowns, provided the sample orientation is kept fixed.
A Fraunhofer type model, where the reduction of $\EJ$ is explained by flux penetration of an extended junction~\cite{Tinkham96}, provides neither qualitative nor quantitative agreement with the data.

Finally, we investigate the coherence properties of the gatemon in $\Bpar$.
At each $\Bpar$ value, the gatemon is placed at a $\Vg$ sweet spot and $\TOne$ and $\TEcho$ are measured [Fig.~\ref{fig:Fig4}(b)].
At low $\Bpar$ (high $\fQ$), $\TOne$ is limited by the Purcell effect ($\TOne^{\rm{P}}$) of the readout resonator.
At high $\Bpar$, the superconducting gap becomes so weak that quasiparticles (assumed to be at an effective temperature of 100~$\mK$) impose a limit~\cite{Riste13}.
By changing $\Vg$, the dependence of $\TOne$ on frequency is measured at different $\Bpar$ [Fig.~\ref{fig:Fig4}(c)].
Heuristically, we describe the structure of the data with a dielectric model.
The data for each value of $\Bpar$ are fitted with $1/\TOne(\fQ)=1/\TOne^{\rm{P}}(\fQ)+2\pi\fQ/\Qdiel(\Bpar)$, where the $\Bpar$-dependent dielectric quality factor $\Qdiel(\Bpar)$ is the only free parameter.
The trend in $\Qdiel(\Bpar)$ shows two plateaus [Fig.~\ref{fig:Fig4}(c) inset]; the value of $\Qdiel$ decreases from $4.6\cdot10^5$ for $\Bpar<6~\mT$ to $2.7\cdot10^5$ for $\Bpar>10~\mT$.
Using an interpolation of $\Qdiel(\Bpar)$ yields an estimate of $\TOne(\Bpar)$ [Fig.~\ref{fig:Fig4}(b)].
A similar trend can be observed for the internal quality factors ($\Qi$) of some resonators in that field range~\cite{SM}.
We could not find an explanation for this trend, but future experiments with resonators designed for field compatibility may help to understand it.
The dip in $\TOne$ at $45~\mT$ is reproducible but hysteretic.
We do not understand its origin.

The dephasing rate $\GEcho$ shows an unexpected dependence on $\Bpar$.
Sweeps of $\Vg$ at fixed values of $\Bpar$ [Fig.~\ref{fig:Fig4}(d)] are performed to extract $\GEcho$ at different $\Vg$ sensitivities.
The minima of the dephasing rates increase with $\Bpar$ and do not quite coincide with the $\Vg$ sweet spot.
The absolute slopes of the dephasing rates can be different on opposite sides of $\partial\fQ/\partial\Vg$.
This is an unusual effect, potentially indicating that the lever arm of the side gate is comparable to the leverarm of the $\Vg$-dependent dominant noise source, which could be confirmed by studying further $\Vg$ sweet spots.
Interpolating the $\GEcho$ traces at each value of $\Bpar$ with two slopes allows extraction of a minimum dephasing rate $\GEchomin$ and an average slope to compute $\SqrtAV$.
A weak decrease of $\SqrtAV$ [Fig.~\ref{fig:Fig4}(d) inset] with $\Bpar$ can be observed.
Direct comparison with Fig.~\ref{fig:Fig3} is not meaningful because the data were obtained in different cooldowns.
Comparing the $\GEcho$ measured at the $\Vg$ sweet spot to $\GEchomin$ highlights that the minimal dephasing rate is obtained at finite $\partial\fQ/\partial\Vg$ [Fig.~\ref{fig:Fig4}(e)].
Finding an explanation of the evolution in $\GEcho(\Bpar)$ [Fig.~\ref{fig:Fig4}(e)] as well as the $\SqrtAV(\Bpar)$ dependence remains for future research.

In conclusion, we have characterized the coherence and noise processes of nanowire transmons at $\Bpar=0$ and explored the performance of a gatemon in $\Bpar$.
The frequency of the gatemon in $\Bpar$ decreases with the reduction of the induced superconducting gap in the nanowire.
Coherence can be observed up to $70~\mT$, limited by the closing of the gap.
Immediate next experiments will focus on finding an explanation for the noise processes taking place at finite $\Bpar$.
By using a persistent current mode for the solenoid providing $\Bpar$, it will also be possible to investigate the spectrum and coherence of flux-tunable split junction devices in $\Bpar$.
This could yield further insight into the microscopic origin of $1/f$ flux noise currently limiting coherence of flux-tunable transmons.
Later, studying the temperature and $\Bpar$ behavior of the observed charge traps will lead to further understanding of their nature.
Using nanowires with a thinner shell (10~$\nm$) covering two to four facets of the wire, while still inducing a hard gap, will allow operation of the qubits up to 0.5~$\Tesla$~\cite{Gazibegovic16}, reaching the relevant range for high-field applications.

\begin{acknowledgements}
We thank A.~Akhmerov, A.~Geresdi, G.~de Lange and M.~de Moor for useful discussions, D.~Thoen for depositing the NbTiN film, and R.~Schouten and J.~Watson for technical assistance. 
We acknowledge funding by Microsoft Corporation Station Q, the Dutch organization for Fundamental Research on Matter (FOM), the Netherlands Organization for Scientific Research (NWO), an ERC Synergy grant, and the Danish National Research Foundation.
\end{acknowledgements}

\clearpage

\renewcommand{\thefigure}{S\arabic{figure}}
\renewcommand{\theequation}{S\arabic{equation}}
\setcounter{figure}{0}
\setcounter{equation}{0}

\section{Supplementary Material for ``Evolution of Nanowire Transmons and Their Quantum Coherence in Magnetic Field''}

\maketitle
This supplement provides experimental details and additional data supporting the claims in the main text.

\section{Experimental Setup}

\begin{figure*}[t]
  \centering
  \includegraphics[width=\linewidth]{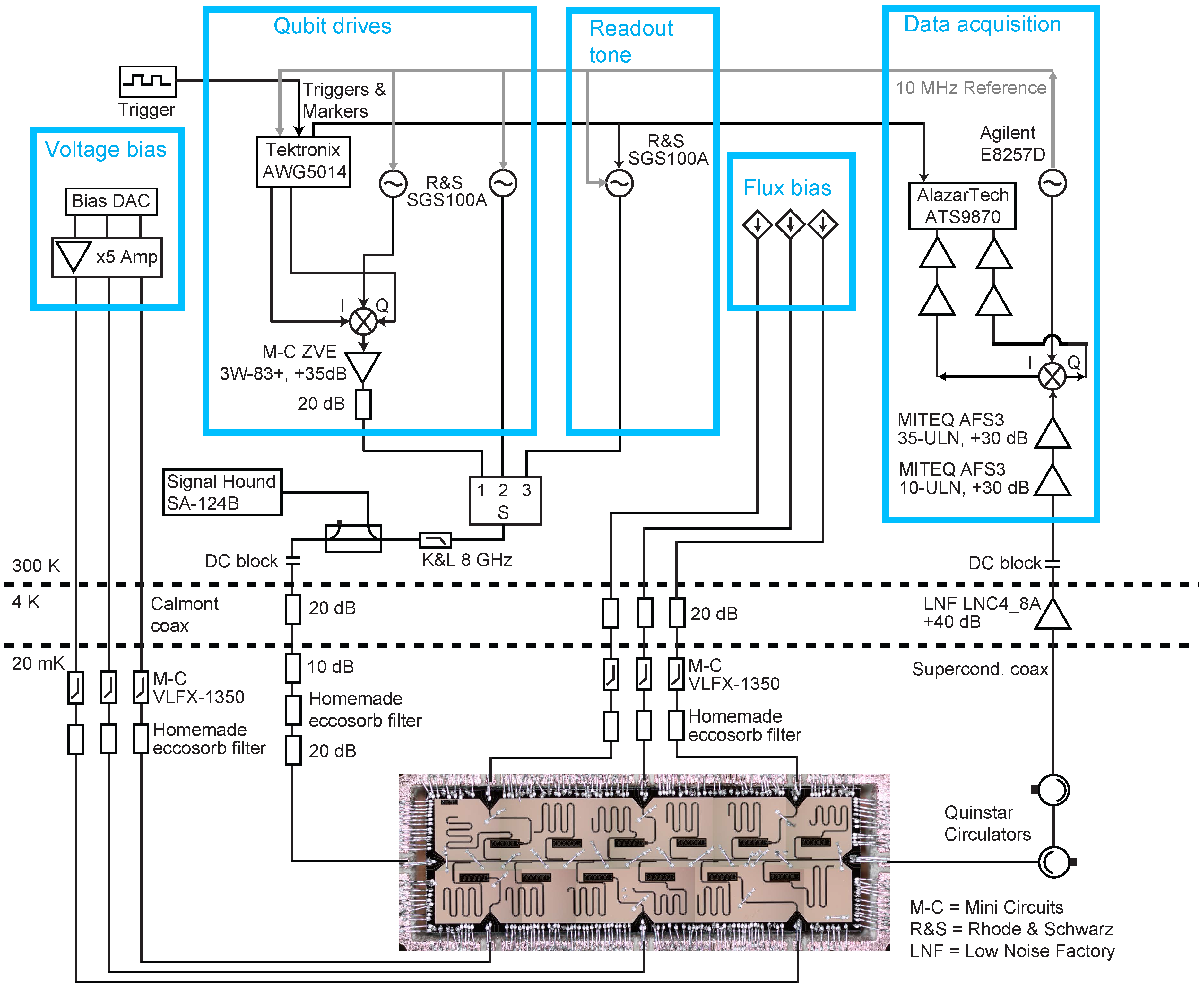}
  \caption{
Optical micrograph of the cQED chip and schematic of the experiment showing microwave and dc connectivity in- and outside the dilution refrigerator.
Silver features crossing the feedline and bias lines are on-chip wire bonds.
  }
  \label{fig:FigS1}
\end{figure*}

Measurements were performed in a variety of experimental conditions, differing in degrees of magnetic and radiation shielding.
The data shown in Figs.~1,~2 and~\ref{fig:FigS3} are taken with the sample in a box that provides radiation and magnetic shielding.
The shielding consists of two layers of Cryophy, a superconducting Al shield and a copper cup coated on the inside with a mixture of silicon carbide and Stycast for infrared shielding~\cite{Barends11}.
The data in other figures are taken with the sample in a copper box, only surrounded by a copper shield coated with the mixture.
Using superconducting shields or passive magnetic shielding was not possible in this situation as this would conflict with the external magnetic field applied.
The coaxial cables carrying voltages, currents and microwave signals are connected to the chip that is mounted on a printed circuit board (PCB) using non-magnetic SMP connectors.
The detailed microwave setup is shown in Fig.~\ref{fig:FigS1}.
Care was taken to only use non-magnetic brass screws in proximity to sample and solenoid.

The magnetic field is generated by a single-axis, cryogen-free, compensated solenoid (American Magnetics, Inc.) with a current-to-field conversion factor 51.6~$\mA/\Tesla$ (max. 2~$\Tesla$), driven by a  Keithley 2200-20-5 programmable power supply.
This solenoid does not have a persistent current switch.

The dc current for flux biasing the split-junction devices is provided by home-built low-noise current sources mounted in a TU Delft IVVI-DAC2 rack.
The voltage to bias the gatemons (provided by DACs of the IVVI rack, amplified with a 5~$\V/\V$ battery-driven amplifier) is low-pass filtered (through Calmont coaxial cables, cutoff frequency 100~$\MHz$, Mini Circuits VLFX 1050 and a home-made, absorptive eccosorb filter) before arriving at the sample.

Microwave tones for qubit control and readout are generated, modulated and combined at room temperature. 
They are coupled to the chip through the common feedline.
Filtering and attenuation at different temperature stages (see Fig.~\ref{fig:FigS1}) suppresses unwanted photon population in the readout resonators.
The readout line wiring is similar to~\cite{Asaad16}.

\section{Fabrication Procedure}

\begin{figure*}[t]
  \centering
  \includegraphics[width=\linewidth]{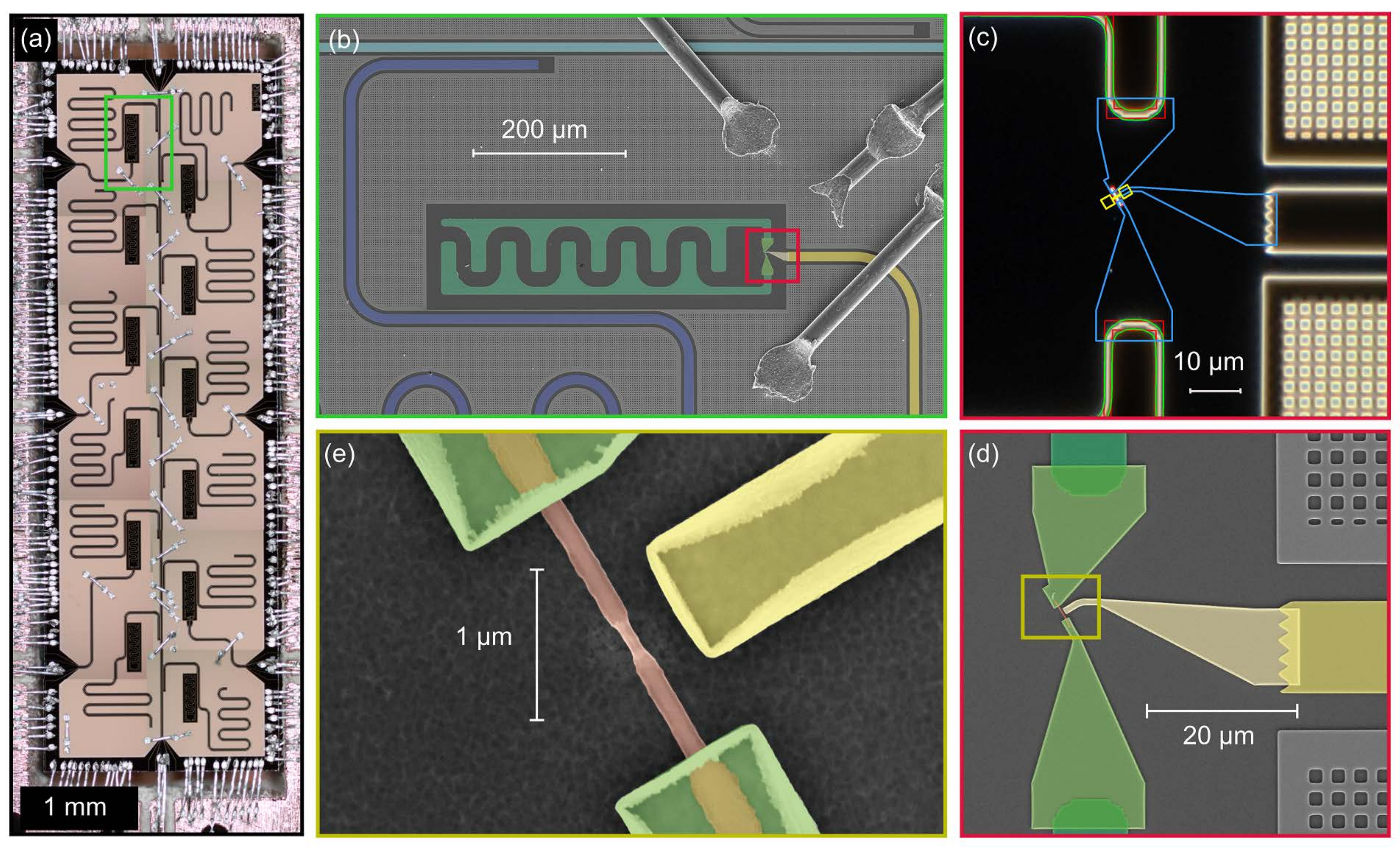}
  \caption{
  Optical and SEM micrographs (added false color) of the sample used for Figs. 1 and 4.
(a) Optical micrograph of the twelve-qubit test chip wirebonded to a PCB.
(b) SEM micrograph zoom-in view on a gatemon qubit (green interdigitated capacitor plates). Readout resonators (blue) connect to a common feedline (light blue). The sidegate (yellow) allows for $\Vg$ tuning of the qubit.
(c) After deterministic deposition of the NW, an optical dark field image is taken. The image recognition software automatically detects contours of wire and leads (light green) and generates an etch mask (bright yellow) and a contact mask (light blue).
(d) SEM micrograph of the same region as (c) after NW etch and contact deposition.
(e) SEM micrograph zoom-in of the region highlighted in (d). The InAs NW (red) is contacted with NbTiN and has a segment of the Al shell etched near the center.
  }
  \label{fig:FigS2}
\end{figure*}

Standard cQED fabrication techniques are used to pattern ground plane, coplanar waveguide structures and qubit capacitor islands~\cite{Bruno15_sup}.
Then, the NWs are deterministically placed in each qubit junction area using a nanomanipulator. 
The InAs NWs have an epitaxially grown Al shell~\cite{Krogstrup14, Chang15_sup} (core diameter 65~$\nm$, shell thickness 30~$\nm$).
Using a PMMA etch mask, a 200~$\nm$ window is etched into the Al shell.
This defines the N section of the SNS junction.
The wet etchant used (Transene D, 12~$\s$ at 50~$\Cel$) is selective enough that no damage to the InAs core can be detected in scanning electron microscope (SEM) micrographs [Fig.~\ref{fig:FigS2}(e)]. 
Some SEM micrographs reveal residues of Al in the junction area of the NWs.
We could not establish a correlation between qubit $\TOne$ or $\TEcho$ and such residues.

To contact the Al shell, a gentle in-situ argon plasma etch is first used to remove the aluminum oxide.
Then, the NbTiN contacts to capacitor plates and sidegates are sputtered.

On-chip Al wirebonds are used to suppress slot-line modes. 
These are added as the sample is also wirebonded to a standard PCB. 

\begin{table*}[t]
  \centering
  \begin{tabular}{| c | c | c | c | c | c | c | c | l |}
    \hline
&&&&&&&&\\[-1em]
     Setup & Qubit & $\fR~[\GHz]$ & $\fQ~[\GHz]$ & $\TOne~[\us]$ & $\TEcho~[\us]$ & $B_c~[\mT]$& $\measuredangle (\Bpar, \mathrm{Wire})$ [$^\circ$]& Comment\\ \hline
 \hline
\multirow{4}{*}{Mag. Shields} & C & 6.08&7.08&15-20&15-20&-&-&2 freqs, 10-15~$\us$ $\Ts$, Single \\ \cline{2-9}
 & H & 6.618 & 6.815 & 7-10 & 6-8 & - &  - &  Single-junction, ungated \\ \cline{2-9}
 & J & 6.832 & 4.9 (SS) & 10-20 & 30 (SS) & - & - & Flux-tunable, Figs.~1,~2 \\ \cline{2-9}
 & K & 6.93 & 5.68 & 2.5-3 & 1.7-2.1 & - & - & Single-junction, ungated \\ \hline
\hline
\multirow{4}{*}{Mag. Shields} & 1 & 6.41 & 7.62 & 5-5 & $\sim$1 & - & - & Gatemon \\ \cline{2-9}
 & 4 & 6.20 & 4.48 & 7-10 & $\sim$4 & - & -  & Single-junction, ungated \\ \cline{2-9}
 & 5  &  6.30 & 5.4-5.8 & 4-7 & 1-2 & - &- & Gatemon, Fig.~\ref{fig:FigS3} \\ \cline{2-9}
 & 6 & 6.42 & 7.62 (SS) & 4-5 & $\sim$1 (SS) & - & - & Flux tunable \\ \hline
\hline
\multirow{5}{*}{Solenoid$\&$Shield} & 1 & 5.88 & 7.06 & $\sim$7 & $\sim$2 & - & 15 & Gatemon \\ \cline{2-9}
 & 4 & 6.21 & 5.49 & $\sim$4.5 & $\sim$4 & 60/59/60 & 50 & Single-junction, ungated \\ \cline{2-9}
 & 5 & 6.31 & 5.08 & 3-4 & 2-3 & 53 & 60 & Gatemon \\ \cline{2-9}
 & 6 & 6.42 & 7.22 & 6-10 & $\sim$8 (SS) & 95/95 & 5 & Flux tunable \\ \cline{2-9}
 & 9 & 6.74 & 4-6.5 & 10-30 & $\sim$20 (SS) & 86/82 & 10 & Gatemon, Figs.~3,~4,~\ref{fig:FigS4} \\ \hline

    \hline
  \end{tabular}
  \caption{Summary of qubit parameters and performance [typical, where indicated on sweet spots (SS)] in different cooldowns. Only working qubits are listed. The two chips investigated have 12 qubits each. Qubits F, G, I, 2, 3, 10 and 11 do not work for fabrication reasons (e.g., displaced NW). It is not clear why the other qubits do not work. Cooldown-to-cooldown reproducibility of $\fQ$, $\TOne$ and $\TEcho$ is limited due to the strong influence of surface absorbents on the NW junction~\cite{Gul15_sup}.}
\end{table*}

\section{Image Recognition Software}

The advantage of top-down fabrication common in cQED is compromised for NW transmons: a bottom-up fabrication approach is required for individual NWs.
Each NW has a different position with respect to the corresponding qubit leads, hence etch mask and contacts (including sidegate) must be individually designed for each qubit. 
To reduce the turnaround time, we wrote software to automatically generate these masks using optical dark field images [compare Fig.~\ref{fig:FigS2}(c)]~\cite{opencv_sup}.

Our image recognition software employs a suite of filtering procedures and feature detection algorithms to reliably design etch and contact masks.
First, the image is low-pass filtered with a Gaussian point spread function.
This reduces the sensitivity to possible dirt in the junction area.
The image is then binarised using Otsu's thresholding method~\cite{Otsu79}.
To further reduce the chance of picking up uninteresting features (such as the holes in the ground plane) and increase the stability of the procedure, a morphological filter combines adjacent areas~\cite{serra1997image}.
The Canny edge detection algorithm finds all contours present in the image.
These are compared to the known shape of the leads to select the best match (green)~\cite{Canny86}.
The scaling, rotation and offset of the image are determined using a Hough transformation and fitting the analytical shape of the leads to the extracted contours (red)~\cite{Ballard1981}.
This allows the definition of a coordinate system.
The NW is then determined as the contour between the leads enclosing the largest area.
The orientation and position of the NW are determined by the smallest rectangle encompassing the NW contour.
This allows correct detection of the NW in $\sim 93$~$\%$ of the cases.
The position and orientation information is used to create the pattern file for the etch windows (yellow), and to place and connect (light blue) contacting regions predefined with respect to leads and wire using a distance minimizing routine.
These contours are used to generate the pattern file for the contact mask.

The performance of the image recognition software is sufficient for our purpose.
The NW width is only 130~$\nm$, well below the diffraction limit (500~$\nm$) and the effective width of the wire in the image ($\sim 1$~$\um$). 
The achieved rms error in sidegate placement is 140~$\nm$.

\section{Gatemon Frequency Switching}

\begin{figure*}[t]
  \centering
  \includegraphics[width=\linewidth]{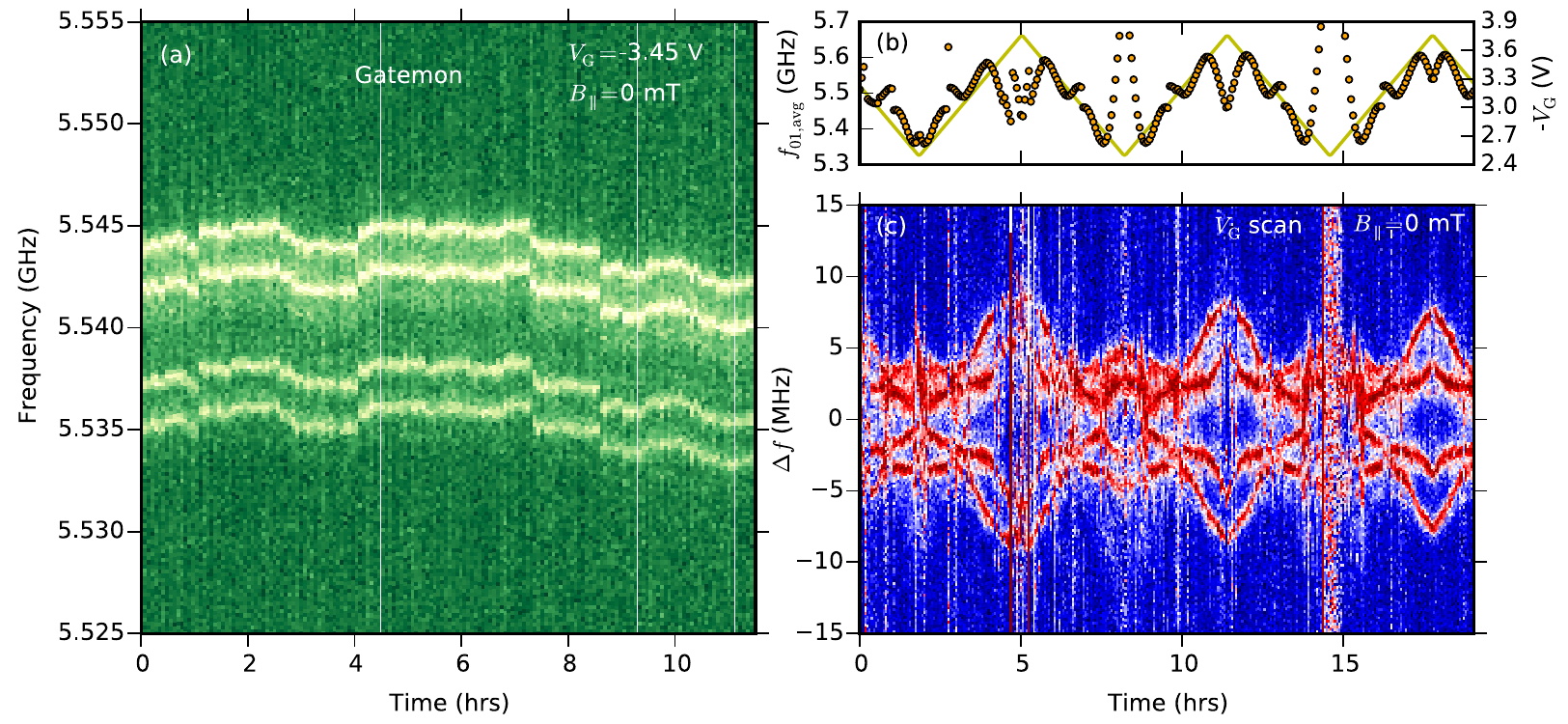}
  \caption{
  Frequency stability of a gatemon.
(a) Repetitions of pulsed qubit spectroscopy scans show four distinct, drifting frequencies between which the qubit switches.
(b) Average $\fQ$ (orange dots) and the triangle profile of $\Vg$ (yellow line) of the scan in (c).
(c) Pulsed qubit spectroscopy while sweeping $\Vg$ in a triangle profile [shown in (b)]. 
Scans are centered around the extracted average $\fQ$. 
The difference between the distinct $\fQ$ values changes with $\Vg$, indicating a sensitivity of the TLSs to the electrostatic environment of the junction.
  }
  \label{fig:FigS3}
\end{figure*}

We have observed the switching of the qubit transition frequency due to strongly coupled TLSs influencing $\EJ$ also in gatemons.
Figure~\ref{fig:FigS3}(a) shows repeated pulsed spectroscopy scans taken at fixed $\Vg=-3.45$~$\V$.
Four distinct, drifting values of $\fQ$ with semi-constant spacings are clearly visible.
A possible explanation of the four frequencies is the strong coupling of two TLSs to $\EJ$.
A background of many weakly coupled TLSs causes the drift in the center frequency.

The difference between the values of $\fQ$ depends on the applied $\Vg$ [Figs.~\ref{fig:FigS3}(b,c)].
Pulsed spectroscopy scans are performed while $\Vg$ is swept up and down.
For each scan, the multiple values of $\fQ$ are extracted and their average is set to $\Delta f=0$~$\MHz$.
The frequency spacing between the peaks changes with $\Vg$.
We therefore interpret the $\Vg$-sensitive TLSs that are influencing $\EJ$ to be charge traps in the vicinity of the NW junction.
The frequencies do not return to the same value upon return to the same $\Vg$.
The drift of the center frequency made it challenging to setup a reliable frequency state measurement (compare Fig.~2). 
Hence, we were not able to estimate the PSD of these TLSs.

\begin{figure*}[t]
  \centering
  \includegraphics[width=\linewidth]{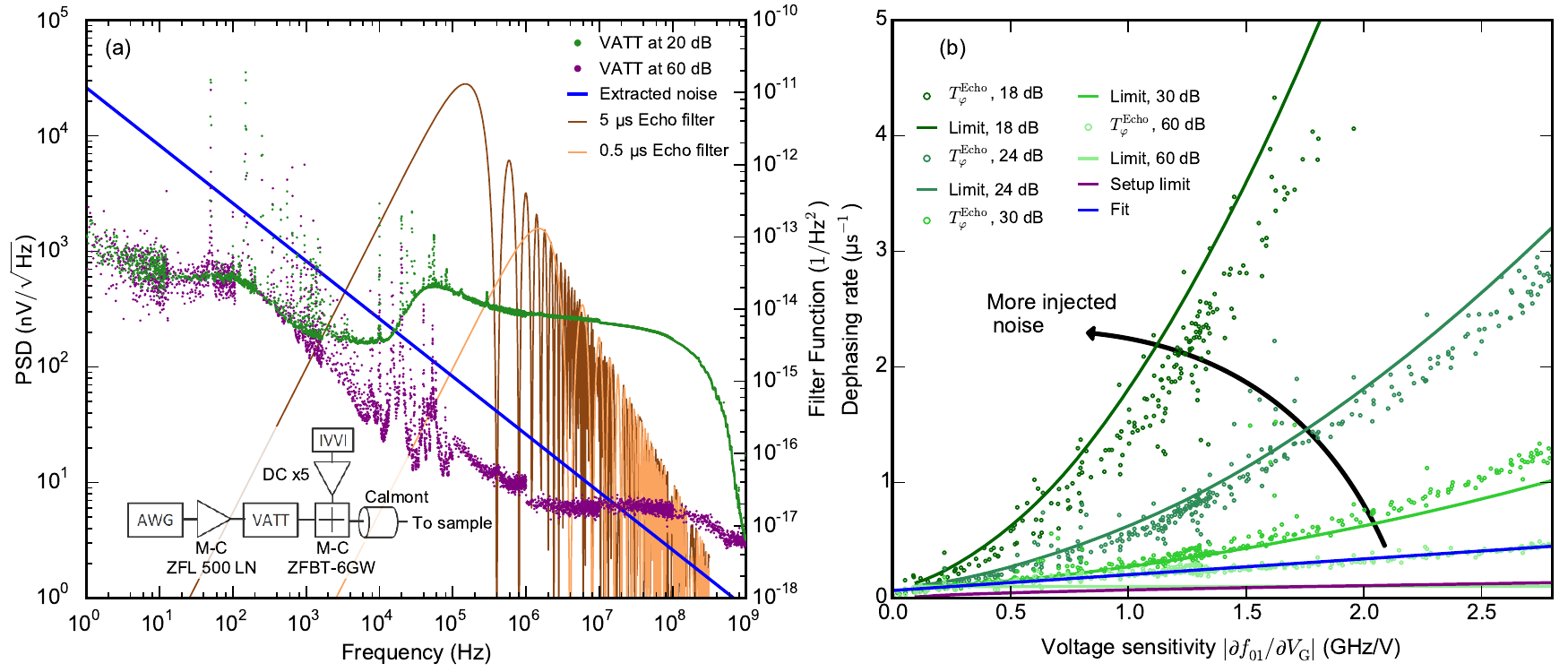}
  \caption{
  Voltage noise injection and coherence time limit.
(a) Measured upper limit to the PSD of the IVVI bias circuit noise (purple) and PSD with additionally injected AWG noise (green), measured after the BiasTee. 
The extracted noise experienced by the gatemon (blue) exceeds the setup noise floor.
Echo filter functions (brown) indicate the frequencies at which the qubit is most sensitive to noise.
Inset: Schematic of the noise injection circuit.
(b) Computed dephasing rate limits (curves) agree well with measured dephasing rates (points) when the injected noise is dominant.
If no noise is injected (purple), another noise source becomes dominant [blue, same fit curve as in Fig.~3(e)].
  }
  \label{fig:FigS4}
\end{figure*}

\section{PSD of the TLS}

The strong coupling of a TLS to the flux-tunable, split junction qubit [Fig.~2(a,b)] allows detailed characterization of the TLS dynamics in real time via Ramsey based time-domain measurements [Fig.~2(c,d)]~\cite{Riste13_sup}.
By monitoring the frequency state of the qubit every $\Delta t =400~\us$ for 6.6~$\s$, we track the TLS state $x_\mathrm{TLS}(t)$ over time.
The PSDs $\mathcal S(f)$ of such traces, given as
\begin{equation}
\mathcal S(f)=\frac{(\Delta t)^2}{T}\left| \sum_{n=1}^{N} x_\mathrm{TLS}(n \cdot\Delta t) e^{-i2\pi f n}\right|^2,
\label{eq:PSD}
\end{equation}
are averaged to get an estimation of the TLS PSD. The TLS PSD can be approximated by an asymmetric random telegraph noise (RTN) model 
\begin{equation}
\mathcal S(f)=\frac{8F^2\Gamma_\uparrow\Gamma_\downarrow}{(\Gamma_\uparrow+\Gamma_\downarrow)((\Gamma_\uparrow+\Gamma_\downarrow)^2+(2\pi f)^2)}+(1-F^2)\Delta t,
\label{eq:RTN}
\end{equation}
where $\Gamma_\uparrow=10.5~\s^{-1}$ and $\Gamma_\downarrow=0.57~\s^{-1}$ are the two switching rates and $F=0.76$ is the detector fidelity [Fig.~2(d)].

Better agreement with the observed data can be achieved by taking the influence of $1/f$ noise into account [Fig.~2(d)].
Given the switching rates $\Gamma_{\uparrow,\downarrow}$, the  noise-free TLS traces are simulated using a Markov chain approach.
Subsequently, $1/f$ frequency noise that is generated by spectrally filtering white noise is superimposed on the TLS traces.
The action of the Ramsey experiment with evolution time $\tau_\mathrm{wait}$ is thresholded to obtain the detector signal $d_\mathrm{TLS}(n\cdot \Delta t)=\mathrm{sign}(\sin(2\pi\cdot f_\mathrm{n}\tau_\mathrm{wait}))$ at the $n$-th time step, at which the frequency of the TLS is $f_\mathrm{n}$.
The detector fidelity (defined as $F=1-\varepsilon_0-\varepsilon_1$, where $\varepsilon_{0,1}$ are the detection error probabilities for the $|0\rangle$ and $|1\rangle$ states) is taken into account by probabilistically flipping the thresholded values.
PSDs of many such traces are calculated using Eq.~(\ref{eq:PSD}) and their average is compared to the experimental PSD.
The experimental parameters of $\fQ$ difference $\Delta f=1.683$~$\MHz$, $\tau_{\mathrm{wait}}=297$~$\ns$ and $\Delta t=400$~$\us$ are used for the simulations.
Switching rates $\Gamma_\uparrow=9.25$~$\persec$ and $\Gamma_\downarrow=0.5$~$\persec$ and fidelity $F=0.76$ agree well with the values found with the asymmetric RTN model [Eq.~(\ref{eq:RTN})].
The additional $1/f$ noise has an amplitude $\sqrt{A_{1/f}}=102$~$\kHzpersqrtHz$ at 1~$\Hz$.
The resulting PSD matches the experimentally obtained PSD better than just an asymmetric RTN curve.
This suggests that $1/f$ noise plays an important role.

\section{Noise PSD extraction from dephasing rates}

The qubits can be used to probe the noise on the control knobs $\lambda$ they are sensitive to.
In the presence of noise in $\lambda$, the $\GEcho$  increases with increasing sensitivity to $\lambda$, $D_\lambda=|\partial \fQ/\partial \lambda|$.
By performing a quadratic fit,
\begin{equation}
\GEcho=aD_\lambda^2+bD_\lambda+c,
\label{eq:quad_fit}
\end{equation}
we can extract the relevant noise parameters~\cite{Martinis03_sup, Yoshihara06_sup, Hutchings17_sup}.
These are the $D_\lambda$-independent offset $c$, a $1/f$ noise contribution linear in $D_\lambda$ and a white noise contribution quadratic in $D_\lambda$.
We quantify the $1/f$ PSD by
\begin{equation}
\sqrt{A_\lambda}=\frac{b}{2\pi\sqrt{\ln(2)}}.
\label{eq:OneFEcho}
\end{equation}
The white noise is quantified by
\begin{equation*}
S_{\mathrm{W},\lambda}=\frac{a}{\pi^2}.
\label{eq:White}
\end{equation*}
Using the qubits as detectors to measure the noise they are subjected to therefore reveals information about the direct environment of the qubits (Figs.~1 and~3).
This is needed to exclude that our control electronics limits the performance of the qubits (Fig.~\ref{fig:FigS4}).

\begin{figure*}[t]
  \centering
  \includegraphics[width=\linewidth]{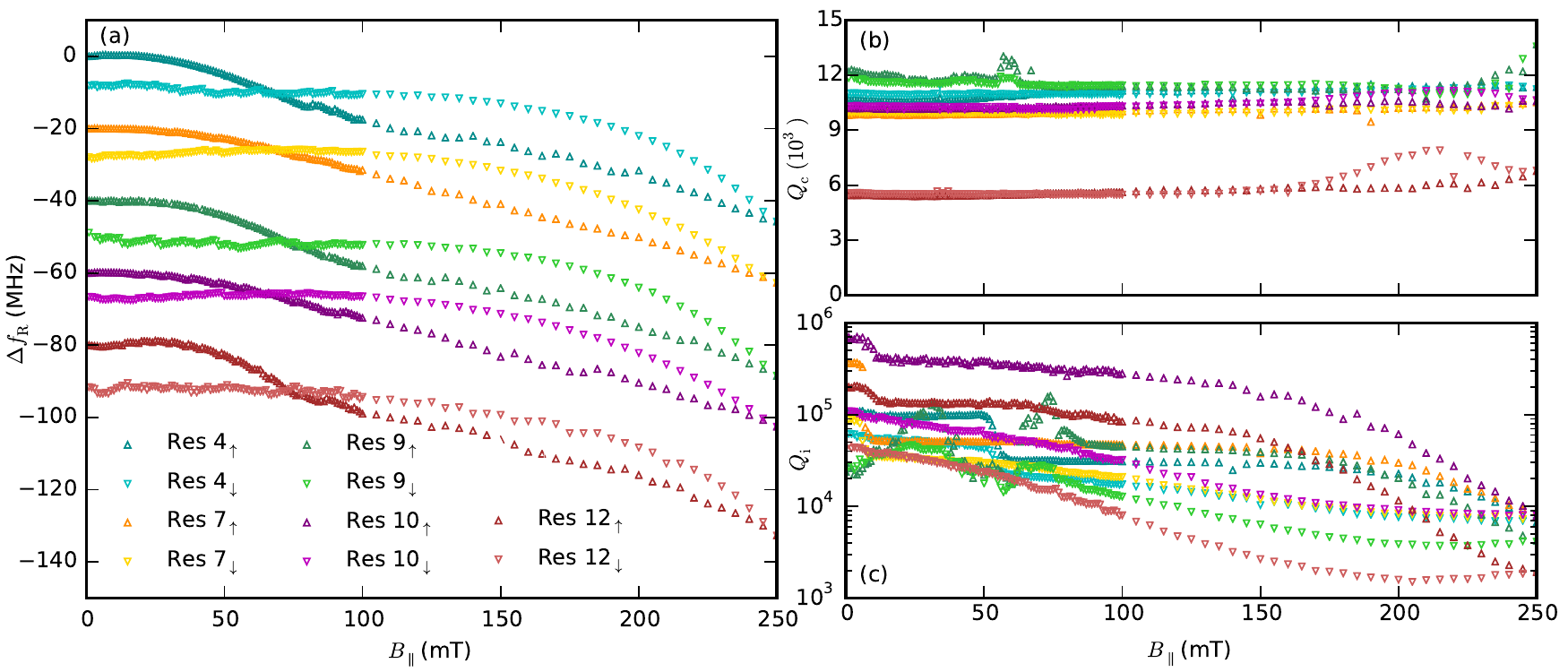}
  \caption{
  Behavior of the resonators when increasing $\Bpar$ from 0 to 250~$\mT$ and back to 0~$\mT$.
(a) Deviation from the fundamental frequency at $\Bpar=0$~$\mT$. Different resonator traces are offset by 20~$\MHz$ each for clarity.
(b) Coupling quality factor $\Qc$ versus $\Bpar$.
(c) Internal quality factor $\Qi$ versus $\Bpar$.
  }
  \label{fig:FigS5}
\end{figure*}

\section{Coherence limitation given a noise PSD}

It is also possible to measure the noise generated by the control electronics and from that calculate a limit on the qubit dephasing time~\cite{Martinis03_sup}.
To do so, the PSD of $\lambda$ is measured at room temperature.
The induced mean-squared phase-noise $\phisqrd$ at a time $t$ is then given as
\begin{equation}
\phisqrd=(2\pi)^2\left(\frac{\partial f_{01}}{\partial \lambda}\right)^2\int_0^{\fQ}\mathcal S_\lambda (f) W(f) \mathrm{d}f,
\label{eq:phisqrd}
\end{equation}
where $W(f)$ is the filter function of the echo sequence used~\cite{Martinis03_sup},
\begin{equation}
W_{\mathrm{SE1}}(f)=\tan^2(\pi f t/2)\frac{\sin^2(\pi f t)}{(\pi f)^2}.
\label{eq:Filter}
\end{equation}
Under the assumption of Gaussian noise, the expected measurement outcome in the computational basis can be expressed as
\begin{align}
\begin{split}
\langle\sigma_z(t)\rangle &= \langle\cos(\phi(t))\rangle = 1-\frac{1}{2}\phisqrd+\frac{3!!}{4!}\phisqrd^2 \pm ...\\
 &=\sum_{n=0}^{\infty} \frac{(-1)^n}{n!}\left(\frac{\phisqrd}{2}\right)^n = \exp \left[-\frac{\langle\phi^2(t)\rangle}{2}\right].
\label{eq:sigmaz}
\end{split}
\end{align}
Inserting the measured PSD and the appropriate filter function [Eq.~(\ref{eq:Filter})] into Eq.~(\ref{eq:phisqrd}) allows us to compute the $1/e$ echo time, using Eq.~(\ref{eq:sigmaz}). 
This provides a tool to calculate an upper limit on the dephasing rates due to the setup.

\section{Voltage noise}

The procedure described above to calculate the dephasing limit is verified on a gatemon, where additional $\Vg$ noise is injected to be the dominating dephasing contribution (Fig.~\ref{fig:FigS4}).
Noise is generated by amplifying the 0-output of a Tektronix AWG 5014 with a Mini-Circuits ZFL 500 LN amplifier. 
Its amplitude is controlled by a Weinschel Aeroflex 8320 variable attenuator (VATT).
The noise is injected to the DC biasing circuit using a Mini-Circuits ZFBT-6GW bias tee [inset Fig.~\ref{fig:FigS4}(a)].
The noise PSDs for the VATT at 60 dB attenuation (no added noise) and at 20 dB (added noise dominates) are measured with a SRS SR770 FFT network analyzer in the range $1-10^5~\Hz$. 
The range between $10^5$ and $10^9$~$\Hz$ is measured with a Rigol DSA 815 spectrum analyzer [Fig.~\ref{fig:FigS4}(a)].
The PSDs are measured after the bias tee, and the transfer function correction of the Calmont coaxial line is applied to the measured spectra.
Note that the noise level measured for the VATT at 60 dB is not discernible from the instrument background.
Hence, this only gives an upper limit to the noise floor.
In Fig.~\ref{fig:FigS4}(b), dephasing rates for different noise levels are plotted and compared to the expected rates given the rescaled injected noise using Eq.~(\ref{eq:sigmaz}).
In the cases where the injected noise is dominant the agreement is good.
A strong deviation becomes apparent when no noise is injected.
This indicates the presence of another noise source.
The extracted noise from that source [Eq.~(\ref{eq:OneFEcho}), Figs.~\ref{fig:FigS4}(a) and 3(f)] exceeds the upper limit on the setup noise floor.

\section{Noise in $\Bpar$}

An accurate estimation of the dephasing time limit imposed by noise in $\Bpar$ is not possible due to the large inductance of the magnet.
Although the current noise of the solenoid biasing circuit was measured, a reliable conversion into an effective noise in $\Bpar$ is not straightforward. 
This is because the (frequency-dependent) conversion function of current to field is not known. 
However, the solenoid acts as a large low-pass filter.
Therefore it is unlikely that this source of noise limits the observed dephasing times.

\section{Resonator performance in $\Bpar$}
The fundamental frequency $\fR$ and the internal quality factor $\Qi$ of the resonators change with applied magnetic field due to the changing kinetic inductance and the induction of vortices in the film. 
An increase in $\Bpar$ means a decrease of the Cooper pair density in the superconducting film, leading to a higher kinetic inductance and thus decreasing $\fR$. 
This effect can be seen in Fig.~\ref{fig:FigS5}(a), showing the deviation from the zero-field $\fR$ of several resonators against $\Bpar$. 
Upon return to zero field, this contribution alone does not lead to a hysteretic effect. 
Hysteretic effects can arise from a net magnetization of the film producing a change in the current distribution of the resonator mode~\cite{Bothner12}.
Vortices induced in the film will experience a Lorentz force due to the current in the resonator, causing them to dissipatively move around, lowering $\Qi$.
The values of $\Qi$ are extracted using real and imaginary part of the feedline transmission~\cite{Bruno15_sup}. 
To speed up measurements, an average intraresonator photon number of~$\sim3000$ was used.
In several of the resonators, a decrease in $\Qi$ between 6 and 10~$\mT$ can be observed.
This is in qualitative agreement with the observed decrease in $\Qdiel(\Bpar)$ in this field range [Fig.~4(c), inset].
Besides this decrease, the performance of the resonators up to 70~$\mT$ allows to perform the experiments presented.


\begin{thebibliography}{46}%
\makeatletter
\providecommand \@ifxundefined [1]{%
 \@ifx{#1\undefined}
}%
\providecommand \@ifnum [1]{%
 \ifnum #1\expandafter \@firstoftwo
 \else \expandafter \@secondoftwo
 \fi
}%
\providecommand \@ifx [1]{%
 \ifx #1\expandafter \@firstoftwo
 \else \expandafter \@secondoftwo
 \fi
}%
\providecommand \natexlab [1]{#1}%
\providecommand \enquote  [1]{``#1''}%
\providecommand \bibnamefont  [1]{#1}%
\providecommand \bibfnamefont [1]{#1}%
\providecommand \citenamefont [1]{#1}%
\providecommand \href@noop [0]{\@secondoftwo}%
\providecommand \href [0]{\begingroup \@sanitize@url \@href}%
\providecommand \@href[1]{\@@startlink{#1}\@@href}%
\providecommand \@@href[1]{\endgroup#1\@@endlink}%
\providecommand \@sanitize@url [0]{\catcode `\\12\catcode `\$12\catcode
  `\&12\catcode `\#12\catcode `\^12\catcode `\_12\catcode `\%12\relax}%
\providecommand \@@startlink[1]{}%
\providecommand \@@endlink[0]{}%
\providecommand \url  [0]{\begingroup\@sanitize@url \@url }%
\providecommand \@url [1]{\endgroup\@href {#1}{\urlprefix }}%
\providecommand \urlprefix  [0]{URL }%
\providecommand \Eprint [0]{\href }%
\providecommand \doibase [0]{http://dx.doi.org/}%
\providecommand \selectlanguage [0]{\@gobble}%
\providecommand \bibinfo  [0]{\@secondoftwo}%
\providecommand \bibfield  [0]{\@secondoftwo}%
\providecommand \translation [1]{[#1]}%
\providecommand \BibitemOpen [0]{}%
\providecommand \bibitemStop [0]{}%
\providecommand \bibitemNoStop [0]{.\EOS\space}%
\providecommand \EOS [0]{\spacefactor3000\relax}%
\providecommand \BibitemShut  [1]{\csname bibitem#1\endcsname}%
\let\auto@bib@innerbib\@empty
\bibitem [{\citenamefont {Blais}\ \emph {et~al.}(2004)\citenamefont {Blais},
  \citenamefont {Huang}, \citenamefont {Wallraff}, \citenamefont {Girvin},\
  and\ \citenamefont {Schoelkopf}}]{Blais04}%
  \BibitemOpen
  \bibfield  {author} {\bibinfo {author} {\bibfnamefont {A.}~\bibnamefont
  {Blais}}, \bibinfo {author} {\bibfnamefont {R.-S.}\ \bibnamefont {Huang}},
  \bibinfo {author} {\bibfnamefont {A.}~\bibnamefont {Wallraff}}, \bibinfo
  {author} {\bibfnamefont {S.~M.}\ \bibnamefont {Girvin}}, \ and\ \bibinfo
  {author} {\bibfnamefont {R.~J.}\ \bibnamefont {Schoelkopf}},\ }\href
  {https://link.aps.org/doi/10.1103/PhysRevA.69.062320} {\bibfield  {journal}
  {\bibinfo  {journal} {Phys. Rev. A}\ }\textbf {\bibinfo {volume} {69}},\
  \bibinfo {pages} {062320} (\bibinfo {year} {2004})}\BibitemShut {NoStop}%
\bibitem [{\citenamefont {Wallraff}\ \emph {et~al.}(2004)\citenamefont
  {Wallraff}, \citenamefont {Schuster}, \citenamefont {Blais}, \citenamefont
  {Frunzio}, \citenamefont {Huang}, \citenamefont {Majer}, \citenamefont
  {Kumar}, \citenamefont {Girvin},\ and\ \citenamefont
  {Schoelkopf}}]{Wallraff04}%
  \BibitemOpen
  \bibfield  {author} {\bibinfo {author} {\bibfnamefont {A.}~\bibnamefont
  {Wallraff}}, \bibinfo {author} {\bibfnamefont {D.~I.}\ \bibnamefont
  {Schuster}}, \bibinfo {author} {\bibfnamefont {A.}~\bibnamefont {Blais}},
  \bibinfo {author} {\bibfnamefont {L.}~\bibnamefont {Frunzio}}, \bibinfo
  {author} {\bibfnamefont {R.-S.}\ \bibnamefont {Huang}}, \bibinfo {author}
  {\bibfnamefont {J.}~\bibnamefont {Majer}}, \bibinfo {author} {\bibfnamefont
  {S.}~\bibnamefont {Kumar}}, \bibinfo {author} {\bibfnamefont {S.~M.}\
  \bibnamefont {Girvin}}, \ and\ \bibinfo {author} {\bibfnamefont {R.~J.}\
  \bibnamefont {Schoelkopf}},\ }\href
  {http://www.nature.com/nature/journal/v431/n7005/abs/nature02851.html}
  {\bibfield  {journal} {\bibinfo  {journal} {Nature}\ }\textbf {\bibinfo
  {volume} {431}},\ \bibinfo {pages} {162} (\bibinfo {year}
  {2004})}\BibitemShut {NoStop}%
\bibitem [{\citenamefont {Josephson}(1962)}]{Josephson1962}%
  \BibitemOpen
  \bibfield  {author} {\bibinfo {author} {\bibfnamefont {B.~D.}\ \bibnamefont
  {Josephson}},\ }\href@noop {} {\bibfield  {journal} {\bibinfo  {journal}
  {Phys. Lett.}\ }\textbf {\bibinfo {volume} {1}},\ \bibinfo {pages} {251}
  (\bibinfo {year} {1962})}\BibitemShut {NoStop}%
\bibitem [{\citenamefont {Nakamura}\ \emph {et~al.}(1999)\citenamefont
  {Nakamura}, \citenamefont {Pashkin},\ and\ \citenamefont
  {Tsai}}]{Nakamura99}%
  \BibitemOpen
  \bibfield  {author} {\bibinfo {author} {\bibfnamefont {Y.}~\bibnamefont
  {Nakamura}}, \bibinfo {author} {\bibfnamefont {Y.}~\bibnamefont {Pashkin}}, \
  and\ \bibinfo {author} {\bibfnamefont {J.}~\bibnamefont {Tsai}},\ }\href
  {http://www.nature.com/nature/journal/v398/n6730/abs/398786a0.html}
  {\bibfield  {journal} {\bibinfo  {journal} {Nature}\ }\textbf {\bibinfo
  {volume} {398}},\ \bibinfo {pages} {786} (\bibinfo {year}
  {1999})}\BibitemShut {NoStop}%
\bibitem [{\citenamefont {Barends}\ \emph {et~al.}(2014)\citenamefont
  {Barends}, \citenamefont {Kelly}, \citenamefont {Megrant}, \citenamefont
  {Veitia}, \citenamefont {Sank}, \citenamefont {Jeffrey}, \citenamefont
  {White}, \citenamefont {Mutus}, \citenamefont {Fowler}, \citenamefont
  {Campbell}, \citenamefont {Chen}, \citenamefont {Chen}, \citenamefont
  {Chiaro}, \citenamefont {Dunsworth}, \citenamefont {Neill}, \citenamefont
  {O'Malley}, \citenamefont {Roushan}, \citenamefont {Vainsencher},
  \citenamefont {Wenner}, \citenamefont {Korotkov}, \citenamefont {Cleland},\
  and\ \citenamefont {Martinis}}]{Barends14}%
  \BibitemOpen
  \bibfield  {author} {\bibinfo {author} {\bibfnamefont {R.}~\bibnamefont
  {Barends}}, \bibinfo {author} {\bibfnamefont {J.}~\bibnamefont {Kelly}},
  \bibinfo {author} {\bibfnamefont {A.}~\bibnamefont {Megrant}}, \bibinfo
  {author} {\bibfnamefont {A.}~\bibnamefont {Veitia}}, \bibinfo {author}
  {\bibfnamefont {D.}~\bibnamefont {Sank}}, \bibinfo {author} {\bibfnamefont
  {E.}~\bibnamefont {Jeffrey}}, \bibinfo {author} {\bibfnamefont {T.~C.}\
  \bibnamefont {White}}, \bibinfo {author} {\bibfnamefont {J.}~\bibnamefont
  {Mutus}}, \bibinfo {author} {\bibfnamefont {A.~G.}\ \bibnamefont {Fowler}},
  \bibinfo {author} {\bibfnamefont {B.}~\bibnamefont {Campbell}}, \bibinfo
  {author} {\bibfnamefont {Y.}~\bibnamefont {Chen}}, \bibinfo {author}
  {\bibfnamefont {Z.}~\bibnamefont {Chen}}, \bibinfo {author} {\bibfnamefont
  {B.}~\bibnamefont {Chiaro}}, \bibinfo {author} {\bibfnamefont
  {A.}~\bibnamefont {Dunsworth}}, \bibinfo {author} {\bibfnamefont
  {C.}~\bibnamefont {Neill}}, \bibinfo {author} {\bibfnamefont
  {P.}~\bibnamefont {O'Malley}}, \bibinfo {author} {\bibfnamefont
  {P.}~\bibnamefont {Roushan}}, \bibinfo {author} {\bibfnamefont
  {A.}~\bibnamefont {Vainsencher}}, \bibinfo {author} {\bibfnamefont
  {J.}~\bibnamefont {Wenner}}, \bibinfo {author} {\bibfnamefont {A.~N.}\
  \bibnamefont {Korotkov}}, \bibinfo {author} {\bibfnamefont {A.~N.}\
  \bibnamefont {Cleland}}, \ and\ \bibinfo {author} {\bibfnamefont {J.~M.}\
  \bibnamefont {Martinis}},\ }\href
  {http://www.nature.com/nature/journal/v508/n7497/abs/nature13171.html}
  {\bibfield  {journal} {\bibinfo  {journal} {Nature}\ }\textbf {\bibinfo
  {volume} {508}},\ \bibinfo {pages} {500} (\bibinfo {year}
  {2014})}\BibitemShut {NoStop}%
\bibitem [{\citenamefont {Steffen}\ \emph {et~al.}(2013)\citenamefont
  {Steffen}, \citenamefont {Salathe}, \citenamefont {Oppliger}, \citenamefont
  {Kurpiers}, \citenamefont {Baur}, \citenamefont {Lang}, \citenamefont
  {Eichler}, \citenamefont {Puebla-Hellmann}, \citenamefont {Fedorov},\ and\
  \citenamefont {Wallraff}}]{Steffen13}%
  \BibitemOpen
  \bibfield  {author} {\bibinfo {author} {\bibfnamefont {L.}~\bibnamefont
  {Steffen}}, \bibinfo {author} {\bibfnamefont {Y.}~\bibnamefont {Salathe}},
  \bibinfo {author} {\bibfnamefont {M.}~\bibnamefont {Oppliger}}, \bibinfo
  {author} {\bibfnamefont {P.}~\bibnamefont {Kurpiers}}, \bibinfo {author}
  {\bibfnamefont {M.}~\bibnamefont {Baur}}, \bibinfo {author} {\bibfnamefont
  {C.}~\bibnamefont {Lang}}, \bibinfo {author} {\bibfnamefont {C.}~\bibnamefont
  {Eichler}}, \bibinfo {author} {\bibfnamefont {G.}~\bibnamefont
  {Puebla-Hellmann}}, \bibinfo {author} {\bibfnamefont {A.}~\bibnamefont
  {Fedorov}}, \ and\ \bibinfo {author} {\bibfnamefont {A.}~\bibnamefont
  {Wallraff}},\ }\href {http://dx.doi.org/10.1038/nature12422} {\bibfield
  {journal} {\bibinfo  {journal} {Nature}\ }\textbf {\bibinfo {volume} {500}},\
  \bibinfo {pages} {319} (\bibinfo {year} {2013})}\BibitemShut {NoStop}%
\bibitem [{\citenamefont {Jerger}\ \emph {et~al.}(2016)\citenamefont {Jerger},
  \citenamefont {Reshitnyk}, \citenamefont {Oppliger}, \citenamefont
  {Poto{\v{c}}nik}, \citenamefont {Mondal}, \citenamefont {Wallraff},
  \citenamefont {Goodenough}, \citenamefont {Wehner}, \citenamefont
  {Juliusson}, \citenamefont {Langford},\ and\ \citenamefont
  {Fedorov}}]{Jerger16}%
  \BibitemOpen
  \bibfield  {author} {\bibinfo {author} {\bibfnamefont {M.}~\bibnamefont
  {Jerger}}, \bibinfo {author} {\bibfnamefont {Y.}~\bibnamefont {Reshitnyk}},
  \bibinfo {author} {\bibfnamefont {M.}~\bibnamefont {Oppliger}}, \bibinfo
  {author} {\bibfnamefont {A.}~\bibnamefont {Poto{\v{c}}nik}}, \bibinfo
  {author} {\bibfnamefont {M.}~\bibnamefont {Mondal}}, \bibinfo {author}
  {\bibfnamefont {A.}~\bibnamefont {Wallraff}}, \bibinfo {author}
  {\bibfnamefont {K.}~\bibnamefont {Goodenough}}, \bibinfo {author}
  {\bibfnamefont {S.}~\bibnamefont {Wehner}}, \bibinfo {author} {\bibfnamefont
  {K.}~\bibnamefont {Juliusson}}, \bibinfo {author} {\bibfnamefont {N.~K.}\
  \bibnamefont {Langford}}, \ and\ \bibinfo {author} {\bibfnamefont
  {A.}~\bibnamefont {Fedorov}},\ }\href@noop {} {\bibfield  {journal} {\bibinfo
   {journal} {Nat.\ Commun.}\ }\textbf {\bibinfo {volume} {7}} (\bibinfo {year}
  {2016})}\BibitemShut {NoStop}%
\bibitem [{\citenamefont {Rist\`{e}}\ \emph {et~al.}(2013)\citenamefont
  {Rist\`{e}}, \citenamefont {Dukalski}, \citenamefont {Watson}, \citenamefont
  {de~Lange}, \citenamefont {Tiggelman}, \citenamefont {Blanter}, \citenamefont
  {Lehnert}, \citenamefont {Schouten},\ and\ \citenamefont
  {DiCarlo}}]{Riste13c}%
  \BibitemOpen
  \bibfield  {author} {\bibinfo {author} {\bibfnamefont {D.}~\bibnamefont
  {Rist\`{e}}}, \bibinfo {author} {\bibfnamefont {M.}~\bibnamefont {Dukalski}},
  \bibinfo {author} {\bibfnamefont {C.~A.}\ \bibnamefont {Watson}}, \bibinfo
  {author} {\bibfnamefont {G.}~\bibnamefont {de~Lange}}, \bibinfo {author}
  {\bibfnamefont {M.~J.}\ \bibnamefont {Tiggelman}}, \bibinfo {author}
  {\bibfnamefont {Y.~M.}\ \bibnamefont {Blanter}}, \bibinfo {author}
  {\bibfnamefont {K.~W.}\ \bibnamefont {Lehnert}}, \bibinfo {author}
  {\bibfnamefont {R.~N.}\ \bibnamefont {Schouten}}, \ and\ \bibinfo {author}
  {\bibfnamefont {L.}~\bibnamefont {DiCarlo}},\ }\href {\doibase
  10.1038/nature12513} {\bibfield  {journal} {\bibinfo  {journal} {Nature}\
  }\textbf {\bibinfo {volume} {502}},\ \bibinfo {pages} {350} (\bibinfo {year}
  {2013})}\BibitemShut {NoStop}%
\bibitem [{\citenamefont {Harris}\ and\ \citenamefont
  {Mapother}(1968)}]{Harris1968}%
  \BibitemOpen
  \bibfield  {author} {\bibinfo {author} {\bibfnamefont {E.~P.}\ \bibnamefont
  {Harris}}\ and\ \bibinfo {author} {\bibfnamefont {D.}~\bibnamefont
  {Mapother}},\ }\href@noop {} {\bibfield  {journal} {\bibinfo  {journal}
  {Phys. Rev.}\ }\textbf {\bibinfo {volume} {165}},\ \bibinfo {pages} {522}
  (\bibinfo {year} {1968})}\BibitemShut {NoStop}%
\bibitem [{\citenamefont {Imamoglu}(2009)}]{Imamoglu09}%
  \BibitemOpen
  \bibfield  {author} {\bibinfo {author} {\bibfnamefont {A.}~\bibnamefont
  {Imamoglu}},\ }\href {\doibase 10.1103/PhysRevLett.102.083602} {\bibfield
  {journal} {\bibinfo  {journal} {Phys. Rev. Lett.}\ }\textbf {\bibinfo
  {volume} {102}},\ \bibinfo {pages} {083602} (\bibinfo {year}
  {2009})}\BibitemShut {NoStop}%
\bibitem [{\citenamefont {Hyart}\ \emph {et~al.}(2013)\citenamefont {Hyart},
  \citenamefont {van Heck}, \citenamefont {Fulga}, \citenamefont {Burrello},
  \citenamefont {Akhmerov},\ and\ \citenamefont {Beenakker}}]{Hyart13}%
  \BibitemOpen
  \bibfield  {author} {\bibinfo {author} {\bibfnamefont {T.}~\bibnamefont
  {Hyart}}, \bibinfo {author} {\bibfnamefont {B.}~\bibnamefont {van Heck}},
  \bibinfo {author} {\bibfnamefont {I.~C.}\ \bibnamefont {Fulga}}, \bibinfo
  {author} {\bibfnamefont {M.}~\bibnamefont {Burrello}}, \bibinfo {author}
  {\bibfnamefont {A.~R.}\ \bibnamefont {Akhmerov}}, \ and\ \bibinfo {author}
  {\bibfnamefont {C.~W.~J.}\ \bibnamefont {Beenakker}},\ }\href {\doibase
  10.1103/PhysRevB.88.035121} {\bibfield  {journal} {\bibinfo  {journal} {Phys.
  Rev. B}\ }\textbf {\bibinfo {volume} {88}},\ \bibinfo {pages} {035121}
  (\bibinfo {year} {2013})}\BibitemShut {NoStop}%
\bibitem [{\citenamefont {Mourik}\ \emph {et~al.}(2012)\citenamefont {Mourik},
  \citenamefont {Zuo}, \citenamefont {Frolov}, \citenamefont {Plissard},
  \citenamefont {Bakkers},\ and\ \citenamefont {Kouwenhoven}}]{Mourik12}%
  \BibitemOpen
  \bibfield  {author} {\bibinfo {author} {\bibfnamefont {V.}~\bibnamefont
  {Mourik}}, \bibinfo {author} {\bibfnamefont {K.}~\bibnamefont {Zuo}},
  \bibinfo {author} {\bibfnamefont {S.~M.}\ \bibnamefont {Frolov}}, \bibinfo
  {author} {\bibfnamefont {S.~R.}\ \bibnamefont {Plissard}}, \bibinfo {author}
  {\bibfnamefont {E.~P. a.~M.}\ \bibnamefont {Bakkers}}, \ and\ \bibinfo
  {author} {\bibfnamefont {L.~P.}\ \bibnamefont {Kouwenhoven}},\ }\href
  {\doibase 10.1126/science.1222360} {\bibfield  {journal} {\bibinfo  {journal}
  {Science}\ }\textbf {\bibinfo {volume} {336}},\ \bibinfo {pages} {1003}
  (\bibinfo {year} {2012})}\BibitemShut {NoStop}%
\bibitem [{\citenamefont {Kosterlitz}\ and\ \citenamefont
  {Thouless}(1973)}]{Kosterlitz1973}%
  \BibitemOpen
  \bibfield  {author} {\bibinfo {author} {\bibfnamefont {J.~M.}\ \bibnamefont
  {Kosterlitz}}\ and\ \bibinfo {author} {\bibfnamefont {D.~J.}\ \bibnamefont
  {Thouless}},\ }\href@noop {} {\bibfield  {journal} {\bibinfo  {journal} {J.
  Phys. C}\ }\textbf {\bibinfo {volume} {6}},\ \bibinfo {pages} {1181}
  (\bibinfo {year} {1973})}\BibitemShut {NoStop}%
\bibitem [{\citenamefont {Andreev}(1964)}]{Andreev1964}%
  \BibitemOpen
  \bibfield  {author} {\bibinfo {author} {\bibfnamefont {A.}~\bibnamefont
  {Andreev}},\ }\href@noop {} {\bibfield  {journal} {\bibinfo  {journal}
  {JETP}\ }\textbf {\bibinfo {volume} {19}},\ \bibinfo {pages} {1228} (\bibinfo
  {year} {1964})}\BibitemShut {NoStop}%
\bibitem [{\citenamefont {Pillet}\ \emph {et~al.}(2010)\citenamefont {Pillet},
  \citenamefont {Quay}, \citenamefont {Morfin}, \citenamefont {Bena},
  \citenamefont {Yeyati},\ and\ \citenamefont {Joyez}}]{Pillet10}%
  \BibitemOpen
  \bibfield  {author} {\bibinfo {author} {\bibfnamefont {J.}~\bibnamefont
  {Pillet}}, \bibinfo {author} {\bibfnamefont {C.}~\bibnamefont {Quay}},
  \bibinfo {author} {\bibfnamefont {P.}~\bibnamefont {Morfin}}, \bibinfo
  {author} {\bibfnamefont {C.}~\bibnamefont {Bena}}, \bibinfo {author}
  {\bibfnamefont {A.~L.}\ \bibnamefont {Yeyati}}, \ and\ \bibinfo {author}
  {\bibfnamefont {P.}~\bibnamefont {Joyez}},\ }\href@noop {} {\bibfield
  {journal} {\bibinfo  {journal} {Nat.\ Phys.}\ }\textbf {\bibinfo {volume}
  {6}},\ \bibinfo {pages} {965} (\bibinfo {year} {2010})}\BibitemShut {NoStop}%
\bibitem [{\citenamefont {van Woerkom}\ \emph {et~al.}(2017)\citenamefont {van
  Woerkom}, \citenamefont {Proutski}, \citenamefont {van Heck}, \citenamefont
  {Bouman}, \citenamefont {V{\"a}yrynen}, \citenamefont {Glazman},
  \citenamefont {Krogstrup}, \citenamefont {Nyg{\aa}rd}, \citenamefont
  {Kouwenhoven},\ and\ \citenamefont {Geresdi}}]{vanWoerkom17}%
  \BibitemOpen
  \bibfield  {author} {\bibinfo {author} {\bibfnamefont {D.~J.}\ \bibnamefont
  {van Woerkom}}, \bibinfo {author} {\bibfnamefont {A.}~\bibnamefont
  {Proutski}}, \bibinfo {author} {\bibfnamefont {B.}~\bibnamefont {van Heck}},
  \bibinfo {author} {\bibfnamefont {D.}~\bibnamefont {Bouman}}, \bibinfo
  {author} {\bibfnamefont {J.~I.}\ \bibnamefont {V{\"a}yrynen}}, \bibinfo
  {author} {\bibfnamefont {L.~I.}\ \bibnamefont {Glazman}}, \bibinfo {author}
  {\bibfnamefont {P.}~\bibnamefont {Krogstrup}}, \bibinfo {author}
  {\bibfnamefont {J.}~\bibnamefont {Nyg{\aa}rd}}, \bibinfo {author}
  {\bibfnamefont {L.~P.}\ \bibnamefont {Kouwenhoven}}, \ and\ \bibinfo {author}
  {\bibfnamefont {A.}~\bibnamefont {Geresdi}},\ }\href@noop {} {\bibfield
  {journal} {\bibinfo  {journal} {Nat.\ Phys.}\ } (\bibinfo {year}
  {2017})}\BibitemShut {NoStop}%
\bibitem [{\citenamefont {Yokoyama}\ \emph {et~al.}(2013)\citenamefont
  {Yokoyama}, \citenamefont {Eto},\ and\ \citenamefont
  {V.~Nazarov}}]{Yokoyama13}%
  \BibitemOpen
  \bibfield  {author} {\bibinfo {author} {\bibfnamefont {T.}~\bibnamefont
  {Yokoyama}}, \bibinfo {author} {\bibfnamefont {M.}~\bibnamefont {Eto}}, \
  and\ \bibinfo {author} {\bibfnamefont {Y.}~\bibnamefont {V.~Nazarov}},\
  }\href@noop {} {\bibfield  {journal} {\bibinfo  {journal} {J. Phys. Soc.
  Jap.}\ }\textbf {\bibinfo {volume} {82}},\ \bibinfo {pages} {054703}
  (\bibinfo {year} {2013})}\BibitemShut {NoStop}%
\bibitem [{\citenamefont {Samkharadze}\ \emph {et~al.}(2016)\citenamefont
  {Samkharadze}, \citenamefont {Bruno}, \citenamefont {Scarlino}, \citenamefont
  {Zheng}, \citenamefont {DiVincenzo}, \citenamefont {DiCarlo},\ and\
  \citenamefont {Vandersypen}}]{Samkharadze16}%
  \BibitemOpen
  \bibfield  {author} {\bibinfo {author} {\bibfnamefont {N.}~\bibnamefont
  {Samkharadze}}, \bibinfo {author} {\bibfnamefont {A.}~\bibnamefont {Bruno}},
  \bibinfo {author} {\bibfnamefont {P.}~\bibnamefont {Scarlino}}, \bibinfo
  {author} {\bibfnamefont {G.}~\bibnamefont {Zheng}}, \bibinfo {author}
  {\bibfnamefont {D.}~\bibnamefont {DiVincenzo}}, \bibinfo {author}
  {\bibfnamefont {L.}~\bibnamefont {DiCarlo}}, \ and\ \bibinfo {author}
  {\bibfnamefont {L.}~\bibnamefont {Vandersypen}},\ }\href@noop {} {\bibfield
  {journal} {\bibinfo  {journal} {Phys. Rev. Appl.}\ }\textbf {\bibinfo
  {volume} {5}},\ \bibinfo {pages} {044004} (\bibinfo {year}
  {2016})}\BibitemShut {NoStop}%
\bibitem [{\citenamefont {Popinciuc}\ \emph {et~al.}(2012)\citenamefont
  {Popinciuc}, \citenamefont {Calado}, \citenamefont {Liu}, \citenamefont
  {Akhmerov}, \citenamefont {Klapwijk},\ and\ \citenamefont
  {Vandersypen}}]{Popinciuc12}%
  \BibitemOpen
  \bibfield  {author} {\bibinfo {author} {\bibfnamefont {M.}~\bibnamefont
  {Popinciuc}}, \bibinfo {author} {\bibfnamefont {V.~E.}\ \bibnamefont
  {Calado}}, \bibinfo {author} {\bibfnamefont {X.~L.}\ \bibnamefont {Liu}},
  \bibinfo {author} {\bibfnamefont {A.~R.}\ \bibnamefont {Akhmerov}}, \bibinfo
  {author} {\bibfnamefont {T.~M.}\ \bibnamefont {Klapwijk}}, \ and\ \bibinfo
  {author} {\bibfnamefont {L.~M.}\ \bibnamefont {Vandersypen}},\ }\href@noop {}
  {\bibfield  {journal} {\bibinfo  {journal} {Phys. Rev. B}\ }\textbf {\bibinfo
  {volume} {85}},\ \bibinfo {pages} {205404} (\bibinfo {year}
  {2012})}\BibitemShut {NoStop}%
\bibitem [{\citenamefont {Doh}\ \emph {et~al.}(2005)\citenamefont {Doh},
  \citenamefont {van Dam}, \citenamefont {Roest}, \citenamefont {Bakkers},
  \citenamefont {Kouwenhoven},\ and\ \citenamefont {De~Franceschi}}]{Doh05}%
  \BibitemOpen
  \bibfield  {author} {\bibinfo {author} {\bibfnamefont {Y.-J.}\ \bibnamefont
  {Doh}}, \bibinfo {author} {\bibfnamefont {J.~A.}\ \bibnamefont {van Dam}},
  \bibinfo {author} {\bibfnamefont {A.~L.}\ \bibnamefont {Roest}}, \bibinfo
  {author} {\bibfnamefont {E.~P.}\ \bibnamefont {Bakkers}}, \bibinfo {author}
  {\bibfnamefont {L.~P.}\ \bibnamefont {Kouwenhoven}}, \ and\ \bibinfo {author}
  {\bibfnamefont {S.}~\bibnamefont {De~Franceschi}},\ }\href@noop {} {\bibfield
   {journal} {\bibinfo  {journal} {Science}\ }\textbf {\bibinfo {volume}
  {309}},\ \bibinfo {pages} {272} (\bibinfo {year} {2005})}\BibitemShut
  {NoStop}%
\bibitem [{\citenamefont {Pallecchi}\ \emph {et~al.}(2008)\citenamefont
  {Pallecchi}, \citenamefont {Gaa{\ss}}, \citenamefont {Ryndyk},\ and\
  \citenamefont {Strunk}}]{Pallecchi08}%
  \BibitemOpen
  \bibfield  {author} {\bibinfo {author} {\bibfnamefont {E.}~\bibnamefont
  {Pallecchi}}, \bibinfo {author} {\bibfnamefont {M.}~\bibnamefont {Gaa{\ss}}},
  \bibinfo {author} {\bibfnamefont {D.}~\bibnamefont {Ryndyk}}, \ and\ \bibinfo
  {author} {\bibfnamefont {C.}~\bibnamefont {Strunk}},\ }\href@noop {}
  {\bibfield  {journal} {\bibinfo  {journal} {Appl. Phys. Lett.}\ }\textbf
  {\bibinfo {volume} {93}},\ \bibinfo {pages} {072501} (\bibinfo {year}
  {2008})}\BibitemShut {NoStop}%
\bibitem [{\citenamefont {Della~Rocca}\ \emph {et~al.}(2007)\citenamefont
  {Della~Rocca}, \citenamefont {Chauvin}, \citenamefont {Huard}, \citenamefont
  {Pothier}, \citenamefont {Esteve},\ and\ \citenamefont {Urbina}}]{Della07}%
  \BibitemOpen
  \bibfield  {author} {\bibinfo {author} {\bibfnamefont {M.}~\bibnamefont
  {Della~Rocca}}, \bibinfo {author} {\bibfnamefont {M.}~\bibnamefont
  {Chauvin}}, \bibinfo {author} {\bibfnamefont {B.}~\bibnamefont {Huard}},
  \bibinfo {author} {\bibfnamefont {H.}~\bibnamefont {Pothier}}, \bibinfo
  {author} {\bibfnamefont {D.}~\bibnamefont {Esteve}}, \ and\ \bibinfo {author}
  {\bibfnamefont {C.}~\bibnamefont {Urbina}},\ }\href@noop {} {\bibfield
  {journal} {\bibinfo  {journal} {Phys. Rev. Lett.}\ }\textbf {\bibinfo
  {volume} {99}},\ \bibinfo {pages} {127005} (\bibinfo {year}
  {2007})}\BibitemShut {NoStop}%
\bibitem [{\citenamefont {Janvier}\ \emph {et~al.}(2015)\citenamefont
  {Janvier}, \citenamefont {Tosi}, \citenamefont {Bretheau}, \citenamefont
  {Girit}, \citenamefont {Stern}, \citenamefont {Bertet}, \citenamefont
  {Joyez}, \citenamefont {Vion}, \citenamefont {Esteve}, \citenamefont
  {Goffman}, \citenamefont {Pothier},\ and\ \citenamefont
  {Urbina}}]{Janvier15}%
  \BibitemOpen
  \bibfield  {author} {\bibinfo {author} {\bibfnamefont {C.}~\bibnamefont
  {Janvier}}, \bibinfo {author} {\bibfnamefont {L.}~\bibnamefont {Tosi}},
  \bibinfo {author} {\bibfnamefont {L.}~\bibnamefont {Bretheau}}, \bibinfo
  {author} {\bibfnamefont {{\c{C}}.}~\bibnamefont {Girit}}, \bibinfo {author}
  {\bibfnamefont {M.}~\bibnamefont {Stern}}, \bibinfo {author} {\bibfnamefont
  {P.}~\bibnamefont {Bertet}}, \bibinfo {author} {\bibfnamefont
  {P.}~\bibnamefont {Joyez}}, \bibinfo {author} {\bibfnamefont
  {D.}~\bibnamefont {Vion}}, \bibinfo {author} {\bibfnamefont {D.}~\bibnamefont
  {Esteve}}, \bibinfo {author} {\bibfnamefont {M.}~\bibnamefont {Goffman}},
  \bibinfo {author} {\bibfnamefont {H.}~\bibnamefont {Pothier}}, \ and\
  \bibinfo {author} {\bibfnamefont {C.}~\bibnamefont {Urbina}},\ }\href@noop {}
  {\bibfield  {journal} {\bibinfo  {journal} {Science}\ }\textbf {\bibinfo
  {volume} {349}},\ \bibinfo {pages} {1199} (\bibinfo {year}
  {2015})}\BibitemShut {NoStop}%
\bibitem [{\citenamefont {de~Lange}\ \emph {et~al.}(2015)\citenamefont
  {de~Lange}, \citenamefont {van Heck}, \citenamefont {Bruno}, \citenamefont
  {van Woerkom}, \citenamefont {Geresdi}, \citenamefont {Plissard},
  \citenamefont {Bakkers}, \citenamefont {Akhmerov},\ and\ \citenamefont
  {DiCarlo}}]{deLange15}%
  \BibitemOpen
  \bibfield  {author} {\bibinfo {author} {\bibfnamefont {G.}~\bibnamefont
  {de~Lange}}, \bibinfo {author} {\bibfnamefont {B.}~\bibnamefont {van Heck}},
  \bibinfo {author} {\bibfnamefont {A.}~\bibnamefont {Bruno}}, \bibinfo
  {author} {\bibfnamefont {D.~J.}\ \bibnamefont {van Woerkom}}, \bibinfo
  {author} {\bibfnamefont {A.}~\bibnamefont {Geresdi}}, \bibinfo {author}
  {\bibfnamefont {S.~R.}\ \bibnamefont {Plissard}}, \bibinfo {author}
  {\bibfnamefont {E.~P. A.~M.}\ \bibnamefont {Bakkers}}, \bibinfo {author}
  {\bibfnamefont {A.~R.}\ \bibnamefont {Akhmerov}}, \ and\ \bibinfo {author}
  {\bibfnamefont {L.}~\bibnamefont {DiCarlo}},\ }\href@noop {} {\bibfield
  {journal} {\bibinfo  {journal} {Phys. Rev. Lett.}\ }\textbf {\bibinfo
  {volume} {115}},\ \bibinfo {pages} {127002} (\bibinfo {year}
  {2015})}\BibitemShut {NoStop}%
\bibitem [{\citenamefont {Larsen}\ \emph {et~al.}(2015)\citenamefont {Larsen},
  \citenamefont {Petersson}, \citenamefont {Kuemmeth}, \citenamefont
  {Jespersen}, \citenamefont {Krogstrup}, \citenamefont {Nyg\aa{}rd},\ and\
  \citenamefont {Marcus}}]{Larsen15}%
  \BibitemOpen
  \bibfield  {author} {\bibinfo {author} {\bibfnamefont {T.~W.}\ \bibnamefont
  {Larsen}}, \bibinfo {author} {\bibfnamefont {K.~D.}\ \bibnamefont
  {Petersson}}, \bibinfo {author} {\bibfnamefont {F.}~\bibnamefont {Kuemmeth}},
  \bibinfo {author} {\bibfnamefont {T.~S.}\ \bibnamefont {Jespersen}}, \bibinfo
  {author} {\bibfnamefont {P.}~\bibnamefont {Krogstrup}}, \bibinfo {author}
  {\bibfnamefont {J.}~\bibnamefont {Nyg\aa{}rd}}, \ and\ \bibinfo {author}
  {\bibfnamefont {C.~M.}\ \bibnamefont {Marcus}},\ }\href@noop {} {\bibfield
  {journal} {\bibinfo  {journal} {Phys. Rev. Lett.}\ }\textbf {\bibinfo
  {volume} {115}},\ \bibinfo {pages} {127001} (\bibinfo {year}
  {2015})}\BibitemShut {NoStop}%
\bibitem [{\citenamefont {Casparis}\ \emph {et~al.}(2016)\citenamefont
  {Casparis}, \citenamefont {Larsen}, \citenamefont {Olsen}, \citenamefont
  {Kuemmeth}, \citenamefont {Krogstrup}, \citenamefont {Nyg\aa{}rd},
  \citenamefont {J.~Petersson},\ and\ \citenamefont {Marcus}}]{Casparis15}%
  \BibitemOpen
  \bibfield  {author} {\bibinfo {author} {\bibfnamefont {L.}~\bibnamefont
  {Casparis}}, \bibinfo {author} {\bibfnamefont {T.~W.}\ \bibnamefont
  {Larsen}}, \bibinfo {author} {\bibfnamefont {M.~S.}\ \bibnamefont {Olsen}},
  \bibinfo {author} {\bibfnamefont {F.}~\bibnamefont {Kuemmeth}}, \bibinfo
  {author} {\bibfnamefont {P.}~\bibnamefont {Krogstrup}}, \bibinfo {author}
  {\bibnamefont {Nyg\aa{}rd}}, \bibinfo {author} {\bibfnamefont {K.~D.}\
  \bibnamefont {J.~Petersson}}, \ and\ \bibinfo {author} {\bibfnamefont
  {C.~M.}\ \bibnamefont {Marcus}},\ }\href@noop {} {\bibfield  {journal}
  {\bibinfo  {journal} {Phys. Rev. Lett.}\ }\textbf {\bibinfo {volume} {116}},\
  \bibinfo {pages} {150505} (\bibinfo {year} {2016})}\BibitemShut {NoStop}%
\bibitem [{\citenamefont {Chuang}\ \emph {et~al.}(2013)\citenamefont {Chuang},
  \citenamefont {Gao}, \citenamefont {Kapadia}, \citenamefont {Ford},
  \citenamefont {Guo},\ and\ \citenamefont {Javey}}]{Chuang13}%
  \BibitemOpen
  \bibfield  {author} {\bibinfo {author} {\bibfnamefont {S.}~\bibnamefont
  {Chuang}}, \bibinfo {author} {\bibfnamefont {Q.}~\bibnamefont {Gao}},
  \bibinfo {author} {\bibfnamefont {R.}~\bibnamefont {Kapadia}}, \bibinfo
  {author} {\bibfnamefont {A.~C.}\ \bibnamefont {Ford}}, \bibinfo {author}
  {\bibfnamefont {J.}~\bibnamefont {Guo}}, \ and\ \bibinfo {author}
  {\bibfnamefont {A.}~\bibnamefont {Javey}},\ }\href@noop {} {\bibfield
  {journal} {\bibinfo  {journal} {Nano Lett.}\ }\textbf {\bibinfo {volume}
  {13}},\ \bibinfo {pages} {555} (\bibinfo {year} {2013})}\BibitemShut
  {NoStop}%
\bibitem [{\citenamefont {Liu}\ \emph {et~al.}(2015)\citenamefont {Liu},
  \citenamefont {Stehlik}, \citenamefont {Eichler}, \citenamefont {Gullans},
  \citenamefont {Taylor},\ and\ \citenamefont {Petta}}]{Liu15}%
  \BibitemOpen
  \bibfield  {author} {\bibinfo {author} {\bibfnamefont {Y.-Y.}\ \bibnamefont
  {Liu}}, \bibinfo {author} {\bibfnamefont {J.}~\bibnamefont {Stehlik}},
  \bibinfo {author} {\bibfnamefont {C.}~\bibnamefont {Eichler}}, \bibinfo
  {author} {\bibfnamefont {M.}~\bibnamefont {Gullans}}, \bibinfo {author}
  {\bibfnamefont {J.~M.}\ \bibnamefont {Taylor}}, \ and\ \bibinfo {author}
  {\bibfnamefont {J.}~\bibnamefont {Petta}},\ }\href@noop {} {\bibfield
  {journal} {\bibinfo  {journal} {Science}\ }\textbf {\bibinfo {volume}
  {347}},\ \bibinfo {pages} {285} (\bibinfo {year} {2015})}\BibitemShut
  {NoStop}%
\bibitem [{\citenamefont {Koch}\ \emph {et~al.}(2007)\citenamefont {Koch},
  \citenamefont {Yu}, \citenamefont {Gambetta}, \citenamefont {Houck},
  \citenamefont {Schuster}, \citenamefont {Majer}, \citenamefont {Blais},
  \citenamefont {Devoret}, \citenamefont {Girvin},\ and\ \citenamefont
  {Schoelkopf}}]{Koch07}%
  \BibitemOpen
  \bibfield  {author} {\bibinfo {author} {\bibfnamefont {J.}~\bibnamefont
  {Koch}}, \bibinfo {author} {\bibfnamefont {T.~M.}\ \bibnamefont {Yu}},
  \bibinfo {author} {\bibfnamefont {J.}~\bibnamefont {Gambetta}}, \bibinfo
  {author} {\bibfnamefont {A.~A.}\ \bibnamefont {Houck}}, \bibinfo {author}
  {\bibfnamefont {D.~I.}\ \bibnamefont {Schuster}}, \bibinfo {author}
  {\bibfnamefont {J.}~\bibnamefont {Majer}}, \bibinfo {author} {\bibfnamefont
  {A.}~\bibnamefont {Blais}}, \bibinfo {author} {\bibfnamefont {M.~H.}\
  \bibnamefont {Devoret}}, \bibinfo {author} {\bibfnamefont {S.~M.}\
  \bibnamefont {Girvin}}, \ and\ \bibinfo {author} {\bibfnamefont {R.~J.}\
  \bibnamefont {Schoelkopf}},\ }\href
  {http://journals.aps.org/pra/abstract/10.1103/PhysRevA.76.042319} {\bibfield
  {journal} {\bibinfo  {journal} {Phys. Rev. A}\ }\textbf {\bibinfo {volume}
  {76}},\ \bibinfo {pages} {042319} (\bibinfo {year} {2007})}\BibitemShut
  {NoStop}%
\bibitem [{\citenamefont {Rist\`e}\ \emph {et~al.}(2013)\citenamefont
  {Rist\`e}, \citenamefont {Bultink}, \citenamefont {Tiggelman}, \citenamefont
  {Schouten}, \citenamefont {Lehnert},\ and\ \citenamefont
  {DiCarlo}}]{Riste13}%
  \BibitemOpen
  \bibfield  {author} {\bibinfo {author} {\bibfnamefont {D.}~\bibnamefont
  {Rist\`e}}, \bibinfo {author} {\bibfnamefont {C.~C.}\ \bibnamefont
  {Bultink}}, \bibinfo {author} {\bibfnamefont {M.~J.}\ \bibnamefont
  {Tiggelman}}, \bibinfo {author} {\bibfnamefont {R.~N.}\ \bibnamefont
  {Schouten}}, \bibinfo {author} {\bibfnamefont {K.~W.}\ \bibnamefont
  {Lehnert}}, \ and\ \bibinfo {author} {\bibfnamefont {L.}~\bibnamefont
  {DiCarlo}},\ }\href {http://www.nature.com/articles/ncomms2936} {\bibfield
  {journal} {\bibinfo  {journal} {Nat.\ Commun.}\ }\textbf {\bibinfo {volume}
  {{4}}},\ \bibinfo {pages} {{1913}} (\bibinfo {year} {{2013}})}\BibitemShut
  {NoStop}%
\bibitem [{\citenamefont {Thoen}\ \emph {et~al.}(2017)\citenamefont {Thoen},
  \citenamefont {Bos}, \citenamefont {Haalebos}, \citenamefont {Klapwijk},
  \citenamefont {Baselmans},\ and\ \citenamefont {Endo}}]{Thoen17}%
  \BibitemOpen
  \bibfield  {author} {\bibinfo {author} {\bibfnamefont {D.~J.}\ \bibnamefont
  {Thoen}}, \bibinfo {author} {\bibfnamefont {B.~G.~C.}\ \bibnamefont {Bos}},
  \bibinfo {author} {\bibfnamefont {E.}~\bibnamefont {Haalebos}}, \bibinfo
  {author} {\bibfnamefont {T.}~\bibnamefont {Klapwijk}}, \bibinfo {author}
  {\bibfnamefont {J.}~\bibnamefont {Baselmans}}, \ and\ \bibinfo {author}
  {\bibfnamefont {A.}~\bibnamefont {Endo}},\ }\href@noop {} {\bibfield
  {journal} {\bibinfo  {journal} {IEEE T.\ Appl.\ Supercon.}\ }\textbf
  {\bibinfo {volume} {27}},\ \bibinfo {pages} {1} (\bibinfo {year}
  {2017})}\BibitemShut {NoStop}%
\bibitem [{\citenamefont {Bos}\ \emph {et~al.}(2017)\citenamefont {Bos},
  \citenamefont {Thoen}, \citenamefont {Haalebos}, \citenamefont {Gimbel},
  \citenamefont {Klapwijk}, \citenamefont {Baselmans},\ and\ \citenamefont
  {Endo}}]{Bos17}%
  \BibitemOpen
  \bibfield  {author} {\bibinfo {author} {\bibfnamefont {B.~G.~C.}\
  \bibnamefont {Bos}}, \bibinfo {author} {\bibfnamefont {D.~J.}\ \bibnamefont
  {Thoen}}, \bibinfo {author} {\bibfnamefont {E.}~\bibnamefont {Haalebos}},
  \bibinfo {author} {\bibfnamefont {P.}~\bibnamefont {Gimbel}}, \bibinfo
  {author} {\bibfnamefont {T.}~\bibnamefont {Klapwijk}}, \bibinfo {author}
  {\bibfnamefont {J.}~\bibnamefont {Baselmans}}, \ and\ \bibinfo {author}
  {\bibfnamefont {A.}~\bibnamefont {Endo}},\ }\href@noop {} {\bibfield
  {journal} {\bibinfo  {journal} {IEEE T.\ Appl.\ Supercon.}\ }\textbf
  {\bibinfo {volume} {27}},\ \bibinfo {pages} {1} (\bibinfo {year}
  {2017})}\BibitemShut {NoStop}%
\bibitem [{\citenamefont {Bruno}\ \emph {et~al.}(2015)\citenamefont {Bruno},
  \citenamefont {de~Lange}, \citenamefont {Asaad}, \citenamefont {van~der
  Enden}, \citenamefont {Langford},\ and\ \citenamefont {DiCarlo}}]{Bruno15}%
  \BibitemOpen
  \bibfield  {author} {\bibinfo {author} {\bibfnamefont {A.}~\bibnamefont
  {Bruno}}, \bibinfo {author} {\bibfnamefont {G.}~\bibnamefont {de~Lange}},
  \bibinfo {author} {\bibfnamefont {S.}~\bibnamefont {Asaad}}, \bibinfo
  {author} {\bibfnamefont {K.~L.}\ \bibnamefont {van~der Enden}}, \bibinfo
  {author} {\bibfnamefont {N.~K.}\ \bibnamefont {Langford}}, \ and\ \bibinfo
  {author} {\bibfnamefont {L.}~\bibnamefont {DiCarlo}},\ }\href
  {http://scitation.aip.org/content/aip/journal/apl/106/18/10.1063/1.4919761}
  {\bibfield  {journal} {\bibinfo  {journal} {Appl. Phys. Lett.}\ }\textbf
  {\bibinfo {volume} {106}},\ \bibinfo {pages} {182601} (\bibinfo {year}
  {2015})}\BibitemShut {NoStop}%
\bibitem [{\citenamefont {Chang}\ \emph {et~al.}(2015)\citenamefont {Chang},
  \citenamefont {Albrecht}, \citenamefont {Jespersen}, \citenamefont
  {Kuemmeth}, \citenamefont {Krogstrup}, \citenamefont {Nyg\r{a}rd},\ and\
  \citenamefont {Marcus}}]{Chang15}%
  \BibitemOpen
  \bibfield  {author} {\bibinfo {author} {\bibfnamefont {W.}~\bibnamefont
  {Chang}}, \bibinfo {author} {\bibfnamefont {S.~M.}\ \bibnamefont {Albrecht}},
  \bibinfo {author} {\bibfnamefont {T.~S.}\ \bibnamefont {Jespersen}}, \bibinfo
  {author} {\bibfnamefont {F.}~\bibnamefont {Kuemmeth}}, \bibinfo {author}
  {\bibfnamefont {P.}~\bibnamefont {Krogstrup}}, \bibinfo {author}
  {\bibfnamefont {J.}~\bibnamefont {Nyg\r{a}rd}}, \ and\ \bibinfo {author}
  {\bibfnamefont {C.~M.}\ \bibnamefont {Marcus}},\ }\href@noop {} {\bibfield
  {journal} {\bibinfo  {journal} {Nat.\ Nanotechnol.}\ }\textbf {\bibinfo
  {volume} {10}} (\bibinfo {year} {2015})}\BibitemShut {NoStop}%
\bibitem [{\citenamefont {Martinis}\ \emph {et~al.}(2009)\citenamefont
  {Martinis}, \citenamefont {Ansmann},\ and\ \citenamefont
  {Aumentado}}]{Martinis09}%
  \BibitemOpen
  \bibfield  {author} {\bibinfo {author} {\bibfnamefont {J.~M.}\ \bibnamefont
  {Martinis}}, \bibinfo {author} {\bibfnamefont {M.}~\bibnamefont {Ansmann}}, \
  and\ \bibinfo {author} {\bibfnamefont {J.}~\bibnamefont {Aumentado}},\ }\href
  {\doibase 10.1103/PhysRevLett.103.097002} {\bibfield  {journal} {\bibinfo
  {journal} {Phys. Rev. Lett.}\ }\textbf {\bibinfo {volume} {103}},\ \bibinfo
  {pages} {097002} (\bibinfo {year} {2009})}\BibitemShut {NoStop}%
\bibitem [{\citenamefont {Wang}\ \emph {et~al.}(2014)\citenamefont {Wang},
  \citenamefont {Gao}, \citenamefont {Pop}, \citenamefont {Vool}, \citenamefont
  {Axline}, \citenamefont {Brecht}, \citenamefont {Heeres}, \citenamefont
  {Frunzio}, \citenamefont {Devoret}, \citenamefont {Catelani}, \citenamefont
  {Glazman},\ and\ \citenamefont {Schoelkopf}}]{Wang14}%
  \BibitemOpen
  \bibfield  {author} {\bibinfo {author} {\bibfnamefont {C.}~\bibnamefont
  {Wang}}, \bibinfo {author} {\bibfnamefont {Y.~Y.}\ \bibnamefont {Gao}},
  \bibinfo {author} {\bibfnamefont {I.~M.}\ \bibnamefont {Pop}}, \bibinfo
  {author} {\bibfnamefont {U.}~\bibnamefont {Vool}}, \bibinfo {author}
  {\bibfnamefont {C.}~\bibnamefont {Axline}}, \bibinfo {author} {\bibfnamefont
  {T.}~\bibnamefont {Brecht}}, \bibinfo {author} {\bibfnamefont {R.~W.}\
  \bibnamefont {Heeres}}, \bibinfo {author} {\bibfnamefont {L.}~\bibnamefont
  {Frunzio}}, \bibinfo {author} {\bibfnamefont {M.~H.}\ \bibnamefont
  {Devoret}}, \bibinfo {author} {\bibfnamefont {G.}~\bibnamefont {Catelani}},
  \bibinfo {author} {\bibfnamefont {L.~I.}\ \bibnamefont {Glazman}}, \ and\
  \bibinfo {author} {\bibfnamefont {R.~J.}\ \bibnamefont {Schoelkopf}},\
  }\href@noop {} {\bibfield  {journal} {\bibinfo  {journal} {Nat.\ Commun.}\
  }\textbf {\bibinfo {volume} {5}},\ \bibinfo {pages} {5836} (\bibinfo {year}
  {2014})}\BibitemShut {NoStop}%
\bibitem [{SM()}]{SM}%
  \BibitemOpen
  \href@noop {} {}\bibinfo {howpublished} {See supplementary
  material.}\BibitemShut {Stop}%
\bibitem [{\citenamefont {Itseez}(2015)}]{opencv}%
  \BibitemOpen
  \bibfield  {author} {\bibinfo {author} {\bibnamefont {Itseez}},\ }\href@noop
  {} {\enquote {\bibinfo {title} {Open source computer vision library},}\
  }\bibinfo {howpublished} {\url{https://github.com/itseez/opencv}} (\bibinfo
  {year} {2015})\BibitemShut {NoStop}%
\bibitem [{\citenamefont {Jerger}\ \emph {et~al.}(2012)\citenamefont {Jerger},
  \citenamefont {Poletto}, \citenamefont {Macha}, \citenamefont {H{\"u}bner},
  \citenamefont {Il'ichev},\ and\ \citenamefont {Ustinov}}]{Jerger12}%
  \BibitemOpen
  \bibfield  {author} {\bibinfo {author} {\bibfnamefont {M.}~\bibnamefont
  {Jerger}}, \bibinfo {author} {\bibfnamefont {S.}~\bibnamefont {Poletto}},
  \bibinfo {author} {\bibfnamefont {P.}~\bibnamefont {Macha}}, \bibinfo
  {author} {\bibfnamefont {U.}~\bibnamefont {H{\"u}bner}}, \bibinfo {author}
  {\bibfnamefont {E.}~\bibnamefont {Il'ichev}}, \ and\ \bibinfo {author}
  {\bibfnamefont {A.~V.}\ \bibnamefont {Ustinov}},\ }\href {\doibase
  http://dx.doi.org/10.1063/1.4739454} {\bibfield  {journal} {\bibinfo
  {journal} {Appl. Phys. Lett.}\ }\textbf {\bibinfo {volume} {101}},\ \bibinfo
  {pages} {042604} (\bibinfo {year} {2012})}\BibitemShut {NoStop}%
\bibitem [{\citenamefont {Motzoi}\ \emph {et~al.}(2009)\citenamefont {Motzoi},
  \citenamefont {Gambetta}, \citenamefont {Rebentrost},\ and\ \citenamefont
  {Wilhelm}}]{Motzoi09}%
  \BibitemOpen
  \bibfield  {author} {\bibinfo {author} {\bibfnamefont {F.}~\bibnamefont
  {Motzoi}}, \bibinfo {author} {\bibfnamefont {J.~M.}\ \bibnamefont
  {Gambetta}}, \bibinfo {author} {\bibfnamefont {P.}~\bibnamefont
  {Rebentrost}}, \ and\ \bibinfo {author} {\bibfnamefont {F.~K.}\ \bibnamefont
  {Wilhelm}},\ }\href@noop {} {\bibfield  {journal} {\bibinfo  {journal} {Phys.
  Rev. Lett.}\ }\textbf {\bibinfo {volume} {103}},\ \bibinfo {pages} {110501}
  (\bibinfo {year} {2009})}\BibitemShut {NoStop}%
\bibitem [{\citenamefont {Martinis}\ \emph {et~al.}(2003)\citenamefont
  {Martinis}, \citenamefont {Nam}, \citenamefont {Aumentado}, \citenamefont
  {Lang},\ and\ \citenamefont {Urbina}}]{Martinis03}%
  \BibitemOpen
  \bibfield  {author} {\bibinfo {author} {\bibfnamefont {J.~M.}\ \bibnamefont
  {Martinis}}, \bibinfo {author} {\bibfnamefont {S.}~\bibnamefont {Nam}},
  \bibinfo {author} {\bibfnamefont {J.}~\bibnamefont {Aumentado}}, \bibinfo
  {author} {\bibfnamefont {K.~M.}\ \bibnamefont {Lang}}, \ and\ \bibinfo
  {author} {\bibfnamefont {C.}~\bibnamefont {Urbina}},\ }\href@noop {}
  {\bibfield  {journal} {\bibinfo  {journal} {Phys. Rev. B}\ }\textbf {\bibinfo
  {volume} {67}},\ \bibinfo {pages} {094510} (\bibinfo {year}
  {2003})}\BibitemShut {NoStop}%
\bibitem [{\citenamefont {Yoshihara}\ \emph {et~al.}(2006)\citenamefont
  {Yoshihara}, \citenamefont {Harrabi}, \citenamefont {Niskanen}, \citenamefont
  {Nakamura},\ and\ \citenamefont {Tsai}}]{Yoshihara06}%
  \BibitemOpen
  \bibfield  {author} {\bibinfo {author} {\bibfnamefont {F.}~\bibnamefont
  {Yoshihara}}, \bibinfo {author} {\bibfnamefont {K.}~\bibnamefont {Harrabi}},
  \bibinfo {author} {\bibfnamefont {A.~O.}\ \bibnamefont {Niskanen}}, \bibinfo
  {author} {\bibfnamefont {Y.}~\bibnamefont {Nakamura}}, \ and\ \bibinfo
  {author} {\bibfnamefont {J.~S.}\ \bibnamefont {Tsai}},\ }\href@noop {}
  {\bibfield  {journal} {\bibinfo  {journal} {Phys. Rev. Lett.}\ }\textbf
  {\bibinfo {volume} {97}},\ \bibinfo {pages} {167001} (\bibinfo {year}
  {2006})}\BibitemShut {NoStop}%
\bibitem [{\citenamefont {Hutchings}\ \emph {et~al.}(2017)\citenamefont
  {Hutchings}, \citenamefont {Hertzberg}, \citenamefont {Liu}, \citenamefont
  {Bronn}, \citenamefont {Keefe}, \citenamefont {Brink}, \citenamefont {Chow},\
  and\ \citenamefont {Plourde}}]{Hutchings17}%
  \BibitemOpen
  \bibfield  {author} {\bibinfo {author} {\bibfnamefont {M.}~\bibnamefont
  {Hutchings}}, \bibinfo {author} {\bibfnamefont {J.~B.}\ \bibnamefont
  {Hertzberg}}, \bibinfo {author} {\bibfnamefont {Y.}~\bibnamefont {Liu}},
  \bibinfo {author} {\bibfnamefont {N.~T.}\ \bibnamefont {Bronn}}, \bibinfo
  {author} {\bibfnamefont {G.~A.}\ \bibnamefont {Keefe}}, \bibinfo {author}
  {\bibfnamefont {M.}~\bibnamefont {Brink}}, \bibinfo {author} {\bibfnamefont
  {J.~M.}\ \bibnamefont {Chow}}, \ and\ \bibinfo {author} {\bibfnamefont
  {B.}~\bibnamefont {Plourde}},\ }\href@noop {} {\bibfield  {journal} {\bibinfo
   {journal} {Phys. Rev. Appl.}\ }\textbf {\bibinfo {volume} {8}},\ \bibinfo
  {pages} {044003} (\bibinfo {year} {2017})}\BibitemShut {NoStop}%
\bibitem [{\citenamefont {G{\"u}l}\ \emph {et~al.}(2015)\citenamefont
  {G{\"u}l}, \citenamefont {Van~Woerkom}, \citenamefont {van Weperen},
  \citenamefont {Car}, \citenamefont {Plissard}, \citenamefont {Bakkers},\ and\
  \citenamefont {Kouwenhoven}}]{Gul15}%
  \BibitemOpen
  \bibfield  {author} {\bibinfo {author} {\bibfnamefont {{\"O}.}~\bibnamefont
  {G{\"u}l}}, \bibinfo {author} {\bibfnamefont {D.~J.}\ \bibnamefont
  {Van~Woerkom}}, \bibinfo {author} {\bibfnamefont {I.}~\bibnamefont {van
  Weperen}}, \bibinfo {author} {\bibfnamefont {D.}~\bibnamefont {Car}},
  \bibinfo {author} {\bibfnamefont {S.~R.}\ \bibnamefont {Plissard}}, \bibinfo
  {author} {\bibfnamefont {E.~P.}\ \bibnamefont {Bakkers}}, \ and\ \bibinfo
  {author} {\bibfnamefont {L.~P.}\ \bibnamefont {Kouwenhoven}},\ }\href@noop {}
  {\bibfield  {journal} {\bibinfo  {journal} {Nanotechnology}\ }\textbf
  {\bibinfo {volume} {26}},\ \bibinfo {pages} {215202} (\bibinfo {year}
  {2015})}\BibitemShut {NoStop}%
\bibitem [{\citenamefont {Tinkham}(1996)}]{Tinkham96}%
  \BibitemOpen
  \bibfield  {author} {\bibinfo {author} {\bibfnamefont {M.}~\bibnamefont
  {Tinkham}},\ }\href@noop {} {\emph {\bibinfo {title} {Introduction to
  Superconductivity}}},\ \bibinfo {edition} {2nd}\ ed.\ (\bibinfo  {publisher}
  {McGraw-Hill},\ \bibinfo {address} {New York},\ \bibinfo {year}
  {1996})\BibitemShut {NoStop}%
\bibitem [{\citenamefont {Gazibegovic}\ \emph {et~al.}(2017)\citenamefont
  {Gazibegovic}, \citenamefont {Car}, \citenamefont {Zhang}, \citenamefont
  {Balk}, \citenamefont {Logan}, \citenamefont {de~Moor}, \citenamefont
  {Cassidy}, \citenamefont {Schmits}, \citenamefont {Xu}, \citenamefont {Wang},
  \citenamefont {Krogstrup}, \citenamefont {Op~het Veld}, \citenamefont {Zuo},
  \citenamefont {Vos}, \citenamefont {Shen}, \citenamefont {Bouman},
  \citenamefont {Shojaei}, \citenamefont {Pennachio}, \citenamefont {Lee},
  \citenamefont {van Veldhoven}, \citenamefont {Koelling}, \citenamefont
  {Verheijen}, \citenamefont {Kouwenhoven}, \citenamefont {Palmstr{\o}m},\ and\
  \citenamefont {Bakkers}}]{Gazibegovic16}%
  \BibitemOpen
  \bibfield  {author} {\bibinfo {author} {\bibfnamefont {S.}~\bibnamefont
  {Gazibegovic}}, \bibinfo {author} {\bibfnamefont {D.}~\bibnamefont {Car}},
  \bibinfo {author} {\bibfnamefont {H.}~\bibnamefont {Zhang}}, \bibinfo
  {author} {\bibfnamefont {S.~C.}\ \bibnamefont {Balk}}, \bibinfo {author}
  {\bibfnamefont {J.~A.}\ \bibnamefont {Logan}}, \bibinfo {author}
  {\bibfnamefont {M.~W.}\ \bibnamefont {de~Moor}}, \bibinfo {author}
  {\bibfnamefont {M.~C.}\ \bibnamefont {Cassidy}}, \bibinfo {author}
  {\bibfnamefont {R.}~\bibnamefont {Schmits}}, \bibinfo {author} {\bibfnamefont
  {D.}~\bibnamefont {Xu}}, \bibinfo {author} {\bibfnamefont {G.}~\bibnamefont
  {Wang}}, \bibinfo {author} {\bibfnamefont {P.}~\bibnamefont {Krogstrup}},
  \bibinfo {author} {\bibfnamefont {R.~L.~M.}\ \bibnamefont {Op~het Veld}},
  \bibinfo {author} {\bibfnamefont {K.}~\bibnamefont {Zuo}}, \bibinfo {author}
  {\bibfnamefont {Y.}~\bibnamefont {Vos}}, \bibinfo {author} {\bibfnamefont
  {J.}~\bibnamefont {Shen}}, \bibinfo {author} {\bibfnamefont {D.}~\bibnamefont
  {Bouman}}, \bibinfo {author} {\bibfnamefont {B.}~\bibnamefont {Shojaei}},
  \bibinfo {author} {\bibfnamefont {D.}~\bibnamefont {Pennachio}}, \bibinfo
  {author} {\bibfnamefont {J.~S.}\ \bibnamefont {Lee}}, \bibinfo {author}
  {\bibfnamefont {P.~J.}\ \bibnamefont {van Veldhoven}}, \bibinfo {author}
  {\bibfnamefont {S.}~\bibnamefont {Koelling}}, \bibinfo {author}
  {\bibfnamefont {M.~A.}\ \bibnamefont {Verheijen}}, \bibinfo {author}
  {\bibfnamefont {L.~P.}\ \bibnamefont {Kouwenhoven}}, \bibinfo {author}
  {\bibfnamefont {C.~J.}\ \bibnamefont {Palmstr{\o}m}}, \ and\ \bibinfo
  {author} {\bibfnamefont {E.~P. A.~M.}\ \bibnamefont {Bakkers}},\ }\href
  {\doibase 10.1038/nature23468} {\bibfield  {journal} {\bibinfo  {journal}
  {Nature}\ }\textbf {\bibinfo {volume} {548}},\ \bibinfo {pages} {434}
  (\bibinfo {year} {2017})}\BibitemShut {NoStop}%
\end{thebibliography}

\begin{thebibliography}{16}%
\makeatletter
\providecommand \@ifxundefined [1]{%
 \@ifx{#1\undefined}
}%
\providecommand \@ifnum [1]{%
 \ifnum #1\expandafter \@firstoftwo
 \else \expandafter \@secondoftwo
 \fi
}%
\providecommand \@ifx [1]{%
 \ifx #1\expandafter \@firstoftwo
 \else \expandafter \@secondoftwo
 \fi
}%
\providecommand \natexlab [1]{#1}%
\providecommand \enquote  [1]{``#1''}%
\providecommand \bibnamefont  [1]{#1}%
\providecommand \bibfnamefont [1]{#1}%
\providecommand \citenamefont [1]{#1}%
\providecommand \href@noop [0]{\@secondoftwo}%
\providecommand \href [0]{\begingroup \@sanitize@url \@href}%
\providecommand \@href[1]{\@@startlink{#1}\@@href}%
\providecommand \@@href[1]{\endgroup#1\@@endlink}%
\providecommand \@sanitize@url [0]{\catcode `\\12\catcode `\$12\catcode
  `\&12\catcode `\#12\catcode `\^12\catcode `\_12\catcode `\%12\relax}%
\providecommand \@@startlink[1]{}%
\providecommand \@@endlink[0]{}%
\providecommand \url  [0]{\begingroup\@sanitize@url \@url }%
\providecommand \@url [1]{\endgroup\@href {#1}{\urlprefix }}%
\providecommand \urlprefix  [0]{URL }%
\providecommand \Eprint [0]{\href }%
\providecommand \doibase [0]{http://dx.doi.org/}%
\providecommand \selectlanguage [0]{\@gobble}%
\providecommand \bibinfo  [0]{\@secondoftwo}%
\providecommand \bibfield  [0]{\@secondoftwo}%
\providecommand \translation [1]{[#1]}%
\providecommand \BibitemOpen [0]{}%
\providecommand \bibitemStop [0]{}%
\providecommand \bibitemNoStop [0]{.\EOS\space}%
\providecommand \EOS [0]{\spacefactor3000\relax}%
\providecommand \BibitemShut  [1]{\csname bibitem#1\endcsname}%
\let\auto@bib@innerbib\@empty
\bibitem [{\citenamefont {Barends}\ \emph {et~al.}(2011)\citenamefont
  {Barends}, \citenamefont {Wenner}, \citenamefont {Lenander}, \citenamefont
  {Chen}, \citenamefont {Bialczak}, \citenamefont {Kelly}, \citenamefont
  {Lucero}, \citenamefont {O'Malley}, \citenamefont {Mariantoni}, \citenamefont
  {Sank}, \citenamefont {Wang}, \citenamefont {White}, \citenamefont {Yin},
  \citenamefont {Zhao}, \citenamefont {Cleland}, \citenamefont {Martinis},\
  and\ \citenamefont {Baselmans}}]{Barends11}%
  \BibitemOpen
  \bibfield  {author} {\bibinfo {author} {\bibfnamefont {R.}~\bibnamefont
  {Barends}}, \bibinfo {author} {\bibfnamefont {J.}~\bibnamefont {Wenner}},
  \bibinfo {author} {\bibfnamefont {M.}~\bibnamefont {Lenander}}, \bibinfo
  {author} {\bibfnamefont {Y.}~\bibnamefont {Chen}}, \bibinfo {author}
  {\bibfnamefont {R.~C.}\ \bibnamefont {Bialczak}}, \bibinfo {author}
  {\bibfnamefont {J.}~\bibnamefont {Kelly}}, \bibinfo {author} {\bibfnamefont
  {E.}~\bibnamefont {Lucero}}, \bibinfo {author} {\bibfnamefont
  {P.}~\bibnamefont {O'Malley}}, \bibinfo {author} {\bibfnamefont
  {M.}~\bibnamefont {Mariantoni}}, \bibinfo {author} {\bibfnamefont
  {D.}~\bibnamefont {Sank}}, \bibinfo {author} {\bibfnamefont {H.}~\bibnamefont
  {Wang}}, \bibinfo {author} {\bibfnamefont {T.~C.}\ \bibnamefont {White}},
  \bibinfo {author} {\bibfnamefont {Y.}~\bibnamefont {Yin}}, \bibinfo {author}
  {\bibfnamefont {J.}~\bibnamefont {Zhao}}, \bibinfo {author} {\bibfnamefont
  {A.~N.}\ \bibnamefont {Cleland}}, \bibinfo {author} {\bibfnamefont {J.~M.}\
  \bibnamefont {Martinis}}, \ and\ \bibinfo {author} {\bibfnamefont {J.~J.~A.}\
  \bibnamefont {Baselmans}},\ }\href
  {http://scitation.aip.org/content/aip/journal/apl/99/11/10.1063/1.3638063}
  {\bibfield  {journal} {\bibinfo  {journal} {Appl. Phys. Lett.}\ }\textbf
  {\bibinfo {volume} {99}},\ \bibinfo {pages} {113507} (\bibinfo {year}
  {2011})}\BibitemShut {NoStop}%
\bibitem [{\citenamefont {Asaad}\ \emph {et~al.}(2016)\citenamefont {Asaad},
  \citenamefont {Dickel}, \citenamefont {Poletto}, \citenamefont {Bruno},
  \citenamefont {Langford}, \citenamefont {Rol}, \citenamefont {Deurloo},\ and\
  \citenamefont {DiCarlo}}]{Asaad16}%
  \BibitemOpen
  \bibfield  {author} {\bibinfo {author} {\bibfnamefont {S.}~\bibnamefont
  {Asaad}}, \bibinfo {author} {\bibfnamefont {C.}~\bibnamefont {Dickel}},
  \bibinfo {author} {\bibfnamefont {S.}~\bibnamefont {Poletto}}, \bibinfo
  {author} {\bibfnamefont {A.}~\bibnamefont {Bruno}}, \bibinfo {author}
  {\bibfnamefont {N.~K.}\ \bibnamefont {Langford}}, \bibinfo {author}
  {\bibfnamefont {M.~A.}\ \bibnamefont {Rol}}, \bibinfo {author} {\bibfnamefont
  {D.}~\bibnamefont {Deurloo}}, \ and\ \bibinfo {author} {\bibfnamefont
  {L.}~\bibnamefont {DiCarlo}},\ }\href
  {https://www.nature.com/articles/npjqi201629} {\bibfield  {journal} {\bibinfo
   {journal} {npj Quantum Inf.}\ }\textbf {\bibinfo {volume} {2}},\ \bibinfo
  {pages} {16029} (\bibinfo {year} {2016})}\BibitemShut {NoStop}%
\bibitem [{\citenamefont {Bruno}\ \emph {et~al.}(2015)\citenamefont {Bruno},
  \citenamefont {de~Lange}, \citenamefont {Asaad}, \citenamefont {van~der
  Enden}, \citenamefont {Langford},\ and\ \citenamefont
  {DiCarlo}}]{Bruno15_sup}%
  \BibitemOpen
  \bibfield  {author} {\bibinfo {author} {\bibfnamefont {A.}~\bibnamefont
  {Bruno}}, \bibinfo {author} {\bibfnamefont {G.}~\bibnamefont {de~Lange}},
  \bibinfo {author} {\bibfnamefont {S.}~\bibnamefont {Asaad}}, \bibinfo
  {author} {\bibfnamefont {K.~L.}\ \bibnamefont {van~der Enden}}, \bibinfo
  {author} {\bibfnamefont {N.~K.}\ \bibnamefont {Langford}}, \ and\ \bibinfo
  {author} {\bibfnamefont {L.}~\bibnamefont {DiCarlo}},\ }\href
  {http://scitation.aip.org/content/aip/journal/apl/106/18/10.1063/1.4919761}
  {\bibfield  {journal} {\bibinfo  {journal} {Appl. Phys. Lett.}\ }\textbf
  {\bibinfo {volume} {106}},\ \bibinfo {pages} {182601} (\bibinfo {year}
  {2015})}\BibitemShut {NoStop}%
\bibitem [{\citenamefont {Krogstrup}\ \emph {et~al.}(2015)\citenamefont
  {Krogstrup}, \citenamefont {Ziino}, \citenamefont {Chang}, \citenamefont
  {Albrecht}, \citenamefont {Madsen}, \citenamefont {Johnson}, \citenamefont
  {Nyg\r{a}rd}, \citenamefont {Marcus},\ and\ \citenamefont
  {Jespersen}}]{Krogstrup14}%
  \BibitemOpen
  \bibfield  {author} {\bibinfo {author} {\bibfnamefont {P.}~\bibnamefont
  {Krogstrup}}, \bibinfo {author} {\bibfnamefont {N.~L.~B.}\ \bibnamefont
  {Ziino}}, \bibinfo {author} {\bibfnamefont {W.}~\bibnamefont {Chang}},
  \bibinfo {author} {\bibfnamefont {S.~M.}\ \bibnamefont {Albrecht}}, \bibinfo
  {author} {\bibfnamefont {M.~H.}\ \bibnamefont {Madsen}}, \bibinfo {author}
  {\bibfnamefont {E.}~\bibnamefont {Johnson}}, \bibinfo {author} {\bibfnamefont
  {J.}~\bibnamefont {Nyg\r{a}rd}}, \bibinfo {author} {\bibfnamefont {C.~M.}\
  \bibnamefont {Marcus}}, \ and\ \bibinfo {author} {\bibfnamefont {T.~S.}\
  \bibnamefont {Jespersen}},\ }\href@noop {} {\bibfield  {journal} {\bibinfo
  {journal} {Nat.\ Mater.}\ }\textbf {\bibinfo {volume} {14}} (\bibinfo {year}
  {2015})}\BibitemShut {NoStop}%
\bibitem [{\citenamefont {Chang}\ \emph {et~al.}(2015)\citenamefont {Chang},
  \citenamefont {Albrecht}, \citenamefont {Jespersen}, \citenamefont
  {Kuemmeth}, \citenamefont {Krogstrup}, \citenamefont {Nyg\r{a}rd},\ and\
  \citenamefont {Marcus}}]{Chang15_sup}%
  \BibitemOpen
  \bibfield  {author} {\bibinfo {author} {\bibfnamefont {W.}~\bibnamefont
  {Chang}}, \bibinfo {author} {\bibfnamefont {S.~M.}\ \bibnamefont {Albrecht}},
  \bibinfo {author} {\bibfnamefont {T.~S.}\ \bibnamefont {Jespersen}}, \bibinfo
  {author} {\bibfnamefont {F.}~\bibnamefont {Kuemmeth}}, \bibinfo {author}
  {\bibfnamefont {P.}~\bibnamefont {Krogstrup}}, \bibinfo {author}
  {\bibfnamefont {J.}~\bibnamefont {Nyg\r{a}rd}}, \ and\ \bibinfo {author}
  {\bibfnamefont {C.~M.}\ \bibnamefont {Marcus}},\ }\href@noop {} {\bibfield
  {journal} {\bibinfo  {journal} {Nat.\ Nanotechnol.}\ }\textbf {\bibinfo
  {volume} {10}} (\bibinfo {year} {2015})}\BibitemShut {NoStop}%
\bibitem [{\citenamefont {G{\"u}l}\ \emph {et~al.}(2015)\citenamefont
  {G{\"u}l}, \citenamefont {Van~Woerkom}, \citenamefont {van Weperen},
  \citenamefont {Car}, \citenamefont {Plissard}, \citenamefont {Bakkers},\ and\
  \citenamefont {Kouwenhoven}}]{Gul15_sup}%
  \BibitemOpen
  \bibfield  {author} {\bibinfo {author} {\bibfnamefont {{\"O}.}~\bibnamefont
  {G{\"u}l}}, \bibinfo {author} {\bibfnamefont {D.~J.}\ \bibnamefont
  {Van~Woerkom}}, \bibinfo {author} {\bibfnamefont {I.}~\bibnamefont {van
  Weperen}}, \bibinfo {author} {\bibfnamefont {D.}~\bibnamefont {Car}},
  \bibinfo {author} {\bibfnamefont {S.~R.}\ \bibnamefont {Plissard}}, \bibinfo
  {author} {\bibfnamefont {E.~P.}\ \bibnamefont {Bakkers}}, \ and\ \bibinfo
  {author} {\bibfnamefont {L.~P.}\ \bibnamefont {Kouwenhoven}},\ }\href@noop {}
  {\bibfield  {journal} {\bibinfo  {journal} {Nanotechnology}\ }\textbf
  {\bibinfo {volume} {26}},\ \bibinfo {pages} {215202} (\bibinfo {year}
  {2015})}\BibitemShut {NoStop}%
\bibitem [{\citenamefont {Itseez}(2015)}]{opencv_sup}%
  \BibitemOpen
  \bibfield  {author} {\bibinfo {author} {\bibnamefont {Itseez}},\ }\href@noop
  {} {\enquote {\bibinfo {title} {Open source computer vision library},}\
  }\bibinfo {howpublished} {\url{https://github.com/itseez/opencv}} (\bibinfo
  {year} {2015})\BibitemShut {NoStop}%
\bibitem [{\citenamefont {Otsu}(1979)}]{Otsu79}%
  \BibitemOpen
  \bibfield  {author} {\bibinfo {author} {\bibfnamefont {N.}~\bibnamefont
  {Otsu}},\ }\href {\doibase 10.1109/TSMC.1979.4310076} {\bibfield  {journal}
  {\bibinfo  {journal} {IEEE T.\ Sys.\ Man and Cyb.}\ }\textbf {\bibinfo
  {volume} {9}},\ \bibinfo {pages} {62 } (\bibinfo {year} {1979})}\BibitemShut
  {NoStop}%
\bibitem [{\citenamefont {Serra}(1997)}]{serra1997image}%
  \BibitemOpen
  \bibfield  {author} {\bibinfo {author} {\bibfnamefont {J.}~\bibnamefont
  {Serra}},\ }\href {https://books.google.nl/books?id=jQwlnAEACAAJ} {\emph
  {\bibinfo {title} {Image Analysis \& Mathematical Morphology}}}\ (\bibinfo
  {publisher} {Academic Press},\ \bibinfo {year} {1997})\BibitemShut {NoStop}%
\bibitem [{\citenamefont {Canny}(1986)}]{Canny86}%
  \BibitemOpen
  \bibfield  {author} {\bibinfo {author} {\bibfnamefont {J.}~\bibnamefont
  {Canny}},\ }\href {\doibase 10.1109/TPAMI.1986.4767851} {\bibfield  {journal}
  {\bibinfo  {journal} {IEEE T.\ PAMI}\ }\textbf {\bibinfo {volume} {8}},\
  \bibinfo {pages} {679 } (\bibinfo {year} {1986})}\BibitemShut {NoStop}%
\bibitem [{\citenamefont {Ballard}(1981)}]{Ballard1981}%
  \BibitemOpen
  \bibfield  {author} {\bibinfo {author} {\bibfnamefont {D.~H.}\ \bibnamefont
  {Ballard}},\ }\href@noop {} {\bibfield  {journal} {\bibinfo  {journal}
  {Pattern recognition}\ }\textbf {\bibinfo {volume} {13}},\ \bibinfo {pages}
  {111} (\bibinfo {year} {1981})}\BibitemShut {NoStop}%
\bibitem [{\citenamefont {Rist\`e}\ \emph {et~al.}(2013)\citenamefont
  {Rist\`e}, \citenamefont {Bultink}, \citenamefont {Tiggelman}, \citenamefont
  {Schouten}, \citenamefont {Lehnert},\ and\ \citenamefont
  {DiCarlo}}]{Riste13_sup}%
  \BibitemOpen
  \bibfield  {author} {\bibinfo {author} {\bibfnamefont {D.}~\bibnamefont
  {Rist\`e}}, \bibinfo {author} {\bibfnamefont {C.~C.}\ \bibnamefont
  {Bultink}}, \bibinfo {author} {\bibfnamefont {M.~J.}\ \bibnamefont
  {Tiggelman}}, \bibinfo {author} {\bibfnamefont {R.~N.}\ \bibnamefont
  {Schouten}}, \bibinfo {author} {\bibfnamefont {K.~W.}\ \bibnamefont
  {Lehnert}}, \ and\ \bibinfo {author} {\bibfnamefont {L.}~\bibnamefont
  {DiCarlo}},\ }\href {http://www.nature.com/articles/ncomms2936} {\bibfield
  {journal} {\bibinfo  {journal} {Nat.\ Commun.}\ }\textbf {\bibinfo {volume}
  {{4}}},\ \bibinfo {pages} {{1913}} (\bibinfo {year} {{2013}})}\BibitemShut
  {NoStop}%
\bibitem [{\citenamefont {Martinis}\ \emph {et~al.}(2003)\citenamefont
  {Martinis}, \citenamefont {Nam}, \citenamefont {Aumentado}, \citenamefont
  {Lang},\ and\ \citenamefont {Urbina}}]{Martinis03_sup}%
  \BibitemOpen
  \bibfield  {author} {\bibinfo {author} {\bibfnamefont {J.~M.}\ \bibnamefont
  {Martinis}}, \bibinfo {author} {\bibfnamefont {S.}~\bibnamefont {Nam}},
  \bibinfo {author} {\bibfnamefont {J.}~\bibnamefont {Aumentado}}, \bibinfo
  {author} {\bibfnamefont {K.~M.}\ \bibnamefont {Lang}}, \ and\ \bibinfo
  {author} {\bibfnamefont {C.}~\bibnamefont {Urbina}},\ }\href@noop {}
  {\bibfield  {journal} {\bibinfo  {journal} {Phys. Rev. B}\ }\textbf {\bibinfo
  {volume} {67}},\ \bibinfo {pages} {094510} (\bibinfo {year}
  {2003})}\BibitemShut {NoStop}%
\bibitem [{\citenamefont {Yoshihara}\ \emph {et~al.}(2006)\citenamefont
  {Yoshihara}, \citenamefont {Harrabi}, \citenamefont {Niskanen}, \citenamefont
  {Nakamura},\ and\ \citenamefont {Tsai}}]{Yoshihara06_sup}%
  \BibitemOpen
  \bibfield  {author} {\bibinfo {author} {\bibfnamefont {F.}~\bibnamefont
  {Yoshihara}}, \bibinfo {author} {\bibfnamefont {K.}~\bibnamefont {Harrabi}},
  \bibinfo {author} {\bibfnamefont {A.~O.}\ \bibnamefont {Niskanen}}, \bibinfo
  {author} {\bibfnamefont {Y.}~\bibnamefont {Nakamura}}, \ and\ \bibinfo
  {author} {\bibfnamefont {J.~S.}\ \bibnamefont {Tsai}},\ }\href@noop {}
  {\bibfield  {journal} {\bibinfo  {journal} {Phys. Rev. Lett.}\ }\textbf
  {\bibinfo {volume} {97}},\ \bibinfo {pages} {167001} (\bibinfo {year}
  {2006})}\BibitemShut {NoStop}%
\bibitem [{\citenamefont {Hutchings}\ \emph {et~al.}(2017)\citenamefont
  {Hutchings}, \citenamefont {Hertzberg}, \citenamefont {Liu}, \citenamefont
  {Bronn}, \citenamefont {Keefe}, \citenamefont {Brink}, \citenamefont {Chow},\
  and\ \citenamefont {Plourde}}]{Hutchings17_sup}%
  \BibitemOpen
  \bibfield  {author} {\bibinfo {author} {\bibfnamefont {M.}~\bibnamefont
  {Hutchings}}, \bibinfo {author} {\bibfnamefont {J.~B.}\ \bibnamefont
  {Hertzberg}}, \bibinfo {author} {\bibfnamefont {Y.}~\bibnamefont {Liu}},
  \bibinfo {author} {\bibfnamefont {N.~T.}\ \bibnamefont {Bronn}}, \bibinfo
  {author} {\bibfnamefont {G.~A.}\ \bibnamefont {Keefe}}, \bibinfo {author}
  {\bibfnamefont {M.}~\bibnamefont {Brink}}, \bibinfo {author} {\bibfnamefont
  {J.~M.}\ \bibnamefont {Chow}}, \ and\ \bibinfo {author} {\bibfnamefont
  {B.}~\bibnamefont {Plourde}},\ }\href@noop {} {\bibfield  {journal} {\bibinfo
   {journal} {Phys. Rev. Appl.}\ }\textbf {\bibinfo {volume} {8}},\ \bibinfo
  {pages} {044003} (\bibinfo {year} {2017})}\BibitemShut {NoStop}%
\bibitem [{\citenamefont {Bothner}\ \emph {et~al.}(2012)\citenamefont
  {Bothner}, \citenamefont {Gaber}, \citenamefont {Kemmler}, \citenamefont
  {Koelle}, \citenamefont {Kleiner}, \citenamefont {W\"unsch},\ and\
  \citenamefont {Siegel}}]{Bothner12}%
  \BibitemOpen
  \bibfield  {author} {\bibinfo {author} {\bibfnamefont {D.}~\bibnamefont
  {Bothner}}, \bibinfo {author} {\bibfnamefont {T.}~\bibnamefont {Gaber}},
  \bibinfo {author} {\bibfnamefont {M.}~\bibnamefont {Kemmler}}, \bibinfo
  {author} {\bibfnamefont {D.}~\bibnamefont {Koelle}}, \bibinfo {author}
  {\bibfnamefont {R.}~\bibnamefont {Kleiner}}, \bibinfo {author} {\bibfnamefont
  {S.}~\bibnamefont {W\"unsch}}, \ and\ \bibinfo {author} {\bibfnamefont
  {M.}~\bibnamefont {Siegel}},\ }\href {\doibase 10.1103/PhysRevB.86.014517}
  {\bibfield  {journal} {\bibinfo  {journal} {Phys. Rev. B}\ }\textbf {\bibinfo
  {volume} {86}},\ \bibinfo {pages} {014517} (\bibinfo {year}
  {2012})}\BibitemShut {NoStop}%
\end{thebibliography}
\end{document}